\documentclass[journal,comsoc, 12pt,onecolumn,draftclsnofoot]{IEEEtran}

\usepackage{textcomp}
\usepackage{verbatim}

\usepackage{tikz}
\usepackage[T1]{fontenc}
\usetikzlibrary{patterns, arrows,backgrounds,fit,tikzmark,positioning,calc,decorations,snakes,shapes,matrix}
\usetikzlibrary{decorations.pathreplacing}
\usetikzlibrary{calc,arrows.meta,positioning}

\usepackage[T1]{fontenc}
\usepackage{lmodern}

\usepackage{float}

\usepackage{cite}

\usepackage{amsfonts}
\usepackage{amsmath,bm, amssymb}
\usepackage{mathtools,hyperref}
\usepackage{amsthm}
\usepackage{mathrsfs}

\usepackage{xcolor}

\usepackage{subcaption} 
\usepackage{nicefrac} 
\usepackage{array}
\usepackage{pgfplots}
\usepackage{pgf,tikz}
\usetikzlibrary{arrows}
\usetikzlibrary{spy}
\usepackage{makecell}

\usepackage[subnum]{cases} 
\usepackage[inline]{enumitem}
\usepackage{lscape}
\usepackage{cleveref}
\usepackage{multicol}
\usepackage{multirow}	

\usepackage{colortbl}
\usepackage{adjustbox}

\theoremstyle{remark}

\newtheorem{lemma}{ {Lemma}}
\newtheorem{remark}{ {Remark}}

\definecolor{RedBlue}{rgb}{0.8,0,0.5}
\definecolor{RedBlueGreen}{rgb}{0.8,0.6,0.5}
\definecolor{YellowOrange}{rgb}{0.4,0.4,0}
\definecolor{OliveGreen}{rgb}{0,0.6,0}
\definecolor{blue-violet}{rgb}{0.54, 0.17, 0.89}
\definecolor{carrotorange}{rgb}{0.93, 0.57, 0.13}
\definecolor{chocolate(traditional)}{rgb}{0.48, 0.25, 0.0}
\definecolor{chartreuse(traditional)}{rgb}{0.87, 1.0, 0.0}
\definecolor{cadmiumyellow}{rgb}{1.0, 0.96, 0.0}
\definecolor{corn}{rgb}{0.98, 0.93, 0.36}
\definecolor{dandelion}{rgb}{0.94, 0.88, 0.19}

\usepackage{footnote}
\makesavenoteenv{tabular}
\makesavenoteenv{table}

\DeclarePairedDelimiter\floor{\lfloor}{\rfloor}

\ifCLASSINFOpdf
\else
\fi
\usepackage{amsmath}
\usepackage[cmintegrals]{newtxmath}
\begin{document}
	
\title{Uplink-Downlink Tradeoff in Secure Distributed Matrix Multiplication}

\author{\IEEEauthorblockN{\small{Jaber~Kakar$^{*}$,~\IEEEmembership{\small Student Member,~IEEE,} Anton~Khristoforov$^{*}$,
        Seyedhamed~Ebadifar$^{*}$,~\IEEEmembership{\small Student Member,~IEEE,} and~Aydin~Sezgin$^{*}$,~\IEEEmembership{\small Senior~Member,~IEEE}\\}
	\IEEEauthorblockA{$^{*}$Institute of Digital Communication Systems,
		Ruhr-Universit{\"a}t Bochum, Germany \\
		Email: \{jaber.kakar, anton.khristoforov, seyedhamed.ebadifar, aydin.sezgin\}@rub.de}
	}\thanks{This work was supported in part by the DFG grant SE 1697/18-1 (COSMOS). This paper will be presented in part at the IEEE International Symposium on Information Theory (ISIT) 2020.}       
}

\markboth{Draft}%
{Shell \MakeLowercase{\textit{et al.}}: Bare Demo of IEEEtran.cls for IEEE Communications Society Journals}

\makeatletter
\newcommand*{\rom}[1]{\expandafter\@slowromancap\romannumeral #1@}
\makeatother

\maketitle

\newcommand{\alert}[1]{\textcolor{black}{#1}}
\newcommand{\alertv}[1]{\textcolor{black}{#1}}

\begin{abstract}
	In secure distributed matrix multiplication (SDMM) the multiplication $\bm A\bm B$ from two private matrices $\bm A$ and $\bm B$ is outsourced by a user to $N$ distributed servers. In $\ell$-SDMM, the goal is to design a joint communication-computation procedure that optimally balances conflicting communication and computation metrics without leaking any information on both $\bm A$ and $\bm B$ to any set of $\ell\leq N$ servers. To this end, the user applies coding with $\tilde{\bm A}_i$ and $\tilde{\bm B}_i$ representing encoded versions of $\bm A$ and $\bm B$ destined to the $i$-th server. Now, SDMM involves multiple tradeoffs. One such tradeoff is the tradeoff between uplink (UL) and downlink (DL) costs. To find a good balance between these two metrics, we propose two schemes which we term USCSA and GSCSA that are based on secure cross subspace alignment (SCSA). We show that there are various scenarios where they outperform existing SDMM schemes from the literature with respect to UL-DL efficiency. Next, we implement schemes from the literature, including USCSA and GSCSA, and test their performance on Amazon EC2. Our numerical results show that USCSA and GSCSA establish a good balance between the time spend on the communication and computation in SDMMs. This is because they combine advantages of polynomial codes, namely low time for the upload of $\left(\tilde{\bm A}_i,\tilde{\bm B}_i\right)_{i=1}^{N}$ and the computation of $\bm O_i=\tilde{\bm A}_i\tilde{\bm B}_i$, with those of SCSA, being a low timing overhead for the download of $\left(\bm O_i\right)_{i=1}^{N}$ and the decoding of $\bm A\bm B$.    
\end{abstract}


\IEEEpeerreviewmaketitle

\section{Introduction}
\label{sec:intro}

Distributed matrix multiplication (DMM) is an important ingredient in many applications, including but not limited to machine learning and object recognition. Recently, coding theory has been applied to enhance the efficiency of DMM. Prominent outcomes of this research thrust are Entangled Polynomial codes \cite{Yu_2018} and PolyDot codes \cite{Dutta_2019}. Although DMM can resolve computational and memory related difficulties, there are security concerns about providing information to external servers. Secure DMM (SDMM) seeks to handle these security concerns through cryptographic, coding and/or information-theoretic means. In the cryptography literature, DMM is applied amongst others in cloud computing \cite{Zhang_2016,Khan_2013} and in the MapReduce framework using partially homomorphic encryption \cite{Bultel_2017}. Recently, it has been shwon that with proper coding, the time needed for SDMM is less than for local matrix multiplication \cite{doliveira2020notes}. In this work, we are interested in SDMM from a coding and information-theoretic perspective.    

\begin{figure}[h]
	\centering
	\begin{tikzpicture}[roundnode/.style={circle, draw=green!60, fill=green!5, very thick, minimum size=7mm}, squarednode/.style={rectangle, draw=red!60, fill=red!5, very thick, minimum size=5mm, rounded corners},scale=0.9]
	
	\node[squarednode,fill=green!30, draw=OliveGreen] (u) at (3, 0) {$\qquad\text{User}\qquad$};
	\node[squarednode] (s1) at (0-0.5, 2.5) {Server $1$};
	\node[squarednode] (s2) at (2, 2.5) {Server $2$};
	\node[scale=2.5] (dots) at (3.75,2.5) {$\ldots$};
	\node[squarednode] (s2) at (5.5, 2.5) {Server $N$};
	
	\draw[->, thick, OliveGreen] (1.75,0.4) -- (-0.25,2.1) node[pos=0.5,sloped,below, black] {\footnotesize $\tilde{\bm A}_{1},\tilde{\bm B}_{1}$};
	\draw[<-, thick, red!60] (2,0.4) -- (0,2.1) node[pos=0.6,sloped,above, black] {\footnotesize $\bm O_{1}$};
	
	\draw[->, thick, OliveGreen] (2.75,0.4) -- (2,2.1) node[pos=0.5,sloped,below, black] {\footnotesize $\tilde{\bm A}_{2},\tilde{\bm B}_{2}$};
	\draw[<-, thick, red!60] (3,0.4) -- (2.25,2.1) node[pos=0.6,sloped,above, black] {\footnotesize $\bm O_{2}$};
	
	\draw[->, thick, OliveGreen] (3.75,0.4) -- (5,2.1) node[pos=0.5,sloped,above, black] {\footnotesize $\tilde{\bm A}_{N},\tilde{\bm B}_{N}$};
	\draw[<-, thick, red!60] (4,0.4) -- (5.25,2.1) node[pos=0.6,sloped,below, black] {\footnotesize $\bm O_{N}$};	
	
	\node (inp) at (-0.1,0) {$\bm A,\bm B$};
	\draw[->, very thick, OliveGreen] (0.5,0) -- (1.45,0) node[pos=0.4,sloped,below, black, scale=0.8] {$\bm f,\bm g$};
	\draw[->, very thick, red!60] (4.55,0) -- (5.5,0) node[pos=0.5,sloped,below, black, scale=0.8] {$d(\cdot)$};;	
	\node (out) at (6,0) {$\bm A\bm B$};
	\end{tikzpicture}
	\caption{\footnotesize System Model of SDMM.}	
	\label{fig:SymMod}
\end{figure}
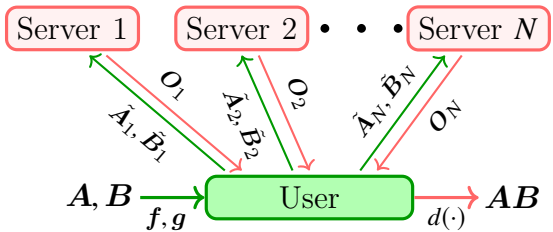

The main problem of SDMM is to effectively retrieve the matrix product $\bm A\bm B$ from $N$ distributed servers without leaking any information on finite field left matrix $\bm A\in\mathbb{F}^{m\times n}$ and right matrix $\bm B\in\mathbb{F}^{n\times p}$ to \emph{any} set of $\ell\leq N$ external servers (cf. Fig \ref{fig:SymMod}). To this end, a user who seeks to determine $\bm A\bm B$, encodes matrices $\bm A$ and $\bm B$ \emph{individually} to encoding matrices $\left\{\tilde{\bm A}_i\right\}_{n=1}^{N}$  and $\left\{\tilde{\bm B}_i\right\}_{i=1}^{N}$. The user then provides the $i$-th server with both $\tilde{\bm A}_i$ and $\tilde{\bm B}_i$ from which the $i$-th server computes $\bm O_i=\tilde{\bm A}_i\tilde{\bm B}_i$. After the download of \emph{any} $Q\leq N$ server observations $\bm O_j$, $j\in\mathcal{Q}\subseteq\left\{1,\ldots,N\right\}$, $|\mathcal{Q}|=Q$, the user shall be able to decode $\bm A\bm B$. In this context, $Q$ is known as the \emph{recovery threshold} (RT). We may measure the efficiency of SDMM through various conflicting metrics.  

Two metrics related to the \emph{communication} efficiency may be the upload ($K_{\text{UL}}$) and download ($K_{\text{DL}}$) communication costs\footnote{We use the words upload and uplink interchangeably. The same applies to download and downlink.}. They measure the ratio of total number of bits uploaded (or downloaded) vs. the number of bits attributed to uploading $(\bm A,\bm B)$ to $N$ servers (or downloading $\bm A\bm B$ from any $Q$ servers). Other metrics are rather concerned with the \emph{computation} complexity. Specifically, the computation complexity can be attributed to both the user and the server. The user computation complexity is comprised of encoding and decoding complexity while the server computation complexity refers to the matrix multiplication $\bm O_i=\tilde{\bm A}_i\tilde{\bm B}_i$. In case of high deviation in server computation times, minimizing the RT $Q$ may be important in order to be robust against particularly slowly computing servers which are frequently termed as stragglers. 

\begin{table}[ht]
	\caption{\small{Related work in the research area of SDMM.}}
	\begin{center}
		\begin{tabular}{|c|c|c|c|}
			\hline
			\multicolumn{4}{|c|}{Metric} \\
			\hline
				\multirow{2}{*}{\makecell{Downlink \\ cost}} & \multirow{2}{*}{\makecell{Recovery \\ threshold}} & 	\multicolumn{2}{c|}{Tradeoff} \\ \cline{3-4}
			& & \makecell{Downlink costs vs. \\ recovery threshold} & \makecell{Uplink vs. \\ downlink costs} \\ \hline 
				\cite{Chang_2018,Kakar_2018,D'Oliveira_2018,Kakar_2019,Jia_2019,Rafael_2019} & \cite{Kakar_2018,Kakar_2019} & \cite{Aliasgari_2019} & \cite{Chang_2019,Jia_2019_PIR} \\ \hline 		
		\end{tabular}
		\label{tab:rel_work_SDMM}			
	\end{center}
\end{table}

The most related work on SDMM based on different performance metrics is given in Table \ref{tab:rel_work_SDMM}. Initially, the research focused on minimizing the download costs for cases where either only one matrix (e.g., $\bm A$) is private \cite{Chang_2018}\footnote{This setting is known as \emph{one-sided} SDMM.} or both matrices $\bm A,\bm B$ ought to be secured \cite{Kakar_2018,Ebadifar_2019,D'Oliveira_2018,Kakar_2019,Jia_2019,Rafael_2019}. For the first scenario, the capacity is fully characterized while for the latter scenario only the asymptotic capacity where the ratio of matrix dimensions satisfy $\nicefrac{n}{\min(m,p)}\rightarrow\infty$ is known. The (asymptotic) capacity-achieving scheme uses a novel interference alignment scheme known as \emph{secure cross subspace alignment} (SCSA) \cite{Jia_2019_TIT}. In this alignment scheme, undesired components are distributed over \emph{multiple} subspace dimensions\footnote{This is in contrast to existing subspace alignment schemes developed for wireless and cache-aided wireless networks \cite{Wang_2014,Kakar_2018_ICC} where one interference component is not attributed to multiple subspaces.}. More recently, rather than optimizing over a single metric, attempts on characterizing the tradeoff between (a) $K_{\text{DL}}$ and RT \cite{Aliasgari_2019} or (b) $K_{\text{DL}}$ and $K_{\text{UL}}$ \cite{Chang_2019,Jia_2019_PIR} have been pursued. In \cite{Chang_2019,Jia_2019_PIR}, a modified system model of SDMM is considered where elements of SDMM are intertwined with private information retrieval (PIR). Specifically, a user is interested in retrieving $\bm A\bm B_{\theta}$, $\theta\in\{1,\ldots,M\}$, privately without revealing any information on $\bm A$ and the realization of $\theta$ to the servers.

In this work, similarly to \cite{Chang_2019,Jia_2019_PIR}, we analyze the tradeoff between $K_{\text{DL}}$ and $K_{\text{UL}}$ for the classical SDMM system model (instead of an intertwined one-sided SDMM-PIR model). To this end, we propose two SCSA-based, uplink cost adjustable schemes which we term USCSA and GSCSA. These two schemes allow us to flexibly balance uplink and downlink costs through respective matrix partitioning and encoding parameters. By choosing USCSA and GSCSA scheme parameters appropriately, we can benefit from advantages of both polynomial and SCSA codes. Note that these schemes resemble an SCSA-scheme developed for the problem of coded distributed \emph{batch} matrix multiplication, which was proposed very recently in an independent work by \emph{Jia et al.} \cite{Jia_2019_CDBMM}. To improve the encoding efficiency of all SDMM schemes, we suggest applying a second-order (SO) polynomial evaluation scheme that exploits the even-odd decomposition of a polynomial $u(x)=u_{\text{even}}\left(x^{2}\right)+xu_{\text{odd}}\left(x^{2}\right)$ evaluated at $x$ and $-x$. To verify our theoretical findings, we test USCSAs and GSCSAs performance against various other schemes with respect to the time needed for encoding, upload, computation, download and decoding through implementation on Amazon EC2 clusters. Our proposed schemes show a particular good balance between the time needed for upload and download. In addition, the simulations show that SO-based polynomial evaluation represents a good encoding solution, particularly for cross subspace alignment (CSA) schemes.     

\textbf{Notation:} Throughout this paper, boldface lower-case and capital letters represent vectors and matrices, respectively. Specifically, for any integer $a$, we define $\bm a_{n}$ to be column vector of dimension $n\times 1$ with all elements being $a$. $\bm A=\mathsf{diag}(\bm x_{n})$ is a diagonal $n\times n$ matrix with the $i$-th diagonal entry $A_{ii}$ corresponding to the $i$-th element $x_{ni}$ of $\bm x_{n}$. We use $\circ$ and $\otimes$ to denote the Hadamard and Kronecker product, respectively. Further, for any two integers $a,b$ with $a\leq b$, we define $[a:b]\triangleq\left\{a,a+1,\ldots,b\right\}$.                           

\section{System Model}
\label{sec:sym_mod}

We consider the problem of secure distributed matrix multiplication (SDMM). In this problem, the user has two \emph{confidential} matrices $\bm A\in\mathbb{F}^{m\times n}$ and $\bm B\in\mathbb{F}^{n\times p}$ with elements drawn from a sufficiently large field $\mathbb{F}$. The goal of the user is to retrieve the matrix product $\bm A\bm B$ by using $N$ servers without revealing the identity of both $\bm A$ and $\bm B$ to the \emph{curious} servers. We assume that any set $\mathcal{L}\subseteq [1:N]$ of $|\mathcal{L}|=\ell\leq N$ collude. The system model of SDMM is shown in Fig. \ref{fig:SymMod}. 

To ensure secure matrix multiplication, the user applies encoding functions $\bm f=(f_1,\ldots,$ $f_N)$ and $\bm g=(g_1,\ldots, g_N)$ to encode matrices $\bm A$ and $\bm B$, respectively. Hereby, $f_i$ and $g_i$ denote functions that encode matrices $\bm A$ and $\bm B$ for the $i$-th server. $\tilde{\bm A}_i$ and $\tilde{\bm B}_i$ are the encoded versions of $\bm A$ and $\bm B$ provided to the $i$-th server; in other words, they are the outputs of encoding functions $f_i$ and $g_i$, i.e.,
\begin{align*}
	\tilde{\bm A}_i=f_i(\bm A),\:\tilde{\bm B}_i=g_i(\bm B).
\end{align*} We assume that every server is \emph{honest}, thus the server observation $\bm O_i$ is a deterministic function of $\tilde{\bm A}_i$ and $\tilde{\bm B}_i$, i.e., $H(\bm O_i|\tilde{\bm A}_i, \tilde{\bm B}_i)=0$. Upon receiving all server observations $\bm O_{[1:N]}=\left(\bm O_1,\ldots, \bm O_N\right)$, the user is able to determine $\bm A\bm B$ by invoking the decoding function $d(\cdot)$, such that $\bm A\bm B=d(\bm O_{[1:N]})$. Information-theoretically, this is equivalent to the \emph{decodability constraint}
\begin{align}
	H\left(\bm A\bm B|\bm O_{[1:N]}\right)=0.\label{eq:decod_constraint}
\end{align} Since servers $j\in\mathcal{L}$ collude and the security has to be preserved, the collection of encoding matrices $\tilde{\bm A}_j$ and $\tilde{\bm B}_j,\forall j\in\mathcal{L}$, denoted by $\tilde{\bm A}_{\mathcal{L}}$ and $\tilde{\bm B}_{\mathcal{L}}$, do not reveal any information on private matrices $\bm A$ and $\bm B$. Thus,
\begin{align}
	I\left(\tilde{\bm A}_{\mathcal{L}},\tilde{\bm B}_{\mathcal{L}};\bm A,\bm B\right)=0,\qquad\forall\mathcal{L}\subseteq[1:N],|\mathcal{L}|=\ell.\label{eq:sec_constraint}
\end{align}
In this paper, we compare the communication cost of SDMM schemes. This cost is comprised of uplink (UL) and downlink (DL) costs. These two costs are defined as follows
\begin{align}
	K_{\text{UL}}&=\frac{\sum_{i=1}^{N}|\tilde{\bm A}_i|+|\tilde{\bm B}_i|}{n(m+p)},\label{eq:KUL}\\
		K_{\text{DL}}&=\frac{\sum_{i=1}^{N}|\bm O_i|}{mp},\label{eq:KDL}	
\end{align} where $|\tilde{\bm A}_i|$, $|\tilde{\bm B}_i|$, $|\bm O_i|$ denote the number of elements 
in $\tilde{\bm A}_i$, $\tilde{\bm B}_i$ and $\bm O_i$, respectively. In \eqref{eq:KUL} and \eqref{eq:KDL}, the denominator corresponds to the number of matrix elements of $(\bm A,\bm B)$ and $\bm A\bm B$, respectively. In the next sections, we elaborate on achievability results on these two metrics. As far as the achievability is concerned, we review the cross subspace alignment scheme \cite{Jia_2019_TIT} applied to SDMM \cite{Kakar_2019,Jia_2019} and then elaborate on our uplink cost adjustable SDMM schemes.

\section{Review: SCSA in SDMM}
\label{cha:rev_SCSA}

Before discussing our scheme that balances uplink and downlink cost, we review SCSA. This scheme is parametrized by the variable $b\in\{0,1\}$. Thus, we write $\text{SCSA}(b)$. In the following, we describe the main ingredients of $\text{SCSA}(b)$, namely matrix partitioning, encoding at the user, matrix multiplication at the servers and the decoding at the user.  

\subsection{Matrix Partitioning} 
Most schemes execute horizontal and vertical matrix partitioning of both $\bm A$ and $\bm B$. To this end, we define the partitioning operator $\mathsf{PART}\left([v_{\bm A},h_{\bm A}],[v_{\bm B},h_{\bm B}]\right)$\footnote{Note that $h_{\bm A}=v_{\bm B}$.}, which breaks matrix $\bm A$ into $v_{\bm A}h_{\bm A}$ equal-size sub-matrices $\bm A_{ij}\in\mathbb{F}^{\nicefrac{m}{v_{\bm A}}\times\:\nicefrac{n}{h_{\bm A}}},\forall i\in[1:v_{\bm A}],j\in[1:h_{\bm A}]$ and matrix $\bm B$ into $v_{\bm B}h_{\bm B}$ sub-matrices $\bm B_{jk}\in\mathbb{F}^{\nicefrac{n}{v_{\bm B}}\times\:\nicefrac{p}{h_{\bm B}}},\forall j\in[1:v_{\bm B}],k\in[1:h_{\bm B}]$ such that
\begin{align*}
	\bm A&=
			\begin{bmatrix}
				\bm A_{11} & \bm A_{12} & \ldots & \bm A_{1h_{\bm A}} \\
					\bm A_{21} & \bm A_{22}		& \ldots & \bm A_{2h_{\bm A}} \\
				\vdots	&	\vdots &	\vdots & \vdots \\
					\bm A_{v_{\bm A}1} & \bm A_{v_{\bm A}2} & \ldots & \bm A_{v_{\bm A}h_{\bm A}}
			\end{bmatrix},\quad
	\bm B=
			\begin{bmatrix}
				\bm B_{11} & \bm B_{12} & \ldots & \bm B_{1h_{\bm B}} \\
					\bm B_{21} & \bm B_{22}		& \ldots & \bm B_{2h_{\bm B}} \\
				\vdots	&	\vdots &	\vdots & \vdots \\
					\bm B_{v_{\bm B}1} & \bm B_{v_{\bm B}2} & \ldots & \bm B_{v_{\bm B}h_{\bm B}}
			\end{bmatrix}.
\end{align*} Note that the $\mathsf{PART}$-operator works under the assumption that $m,n$ and $p$ are multiple of $v_{\bm A},h_{\bm A}$ and $h_{\bm B}$, respectively. In $\text{SCSA}(b)$, $b\in\{0,1\}$, the user either applies \emph{only} a partitioning 
\begin{itemize}
	\item of matrix $\bm A$ in the form $\mathsf{PART}([r,1],[1,1])$ if $b=0$,
	\item or of matrix $\bm B$ in the form $\mathsf{PART}([1,1],[1,r])$ if $b=1$
\end{itemize} with $r=N-2\ell$. In the sequel of this chapter, we describe latter case $(b=1)$. For this case
\begin{align*}
	\bm A\bm B=
		\begin{bmatrix}
			\bm A \bm B_1 & \bm A \bm B_2 & \hdots & \bm A \bm B_{r}
		\end{bmatrix}. 
\end{align*}

\subsection{Encoding at the User}

Based on the above partition, the user encodes matrix $\bm A$ and each sub-matrix $\bm B_j$ (destined to the $i$-th server) for $\text{SCSA}(1)$ individually according to:
\begin{align}
	\tilde{\bm A}_{i}^{(j)}&=\frac{1}{j+\alpha_i}\bigg(\bm A+\sum_{k=1}^{\ell}(j+\alpha_i)^{k}\bm Z_{jk}\bigg),\label{eq:cross_aligment_enc_A}\\
		\tilde{\bm B}_{i}^{(j)}&=\bm B_j+\sum_{k=1}^{\ell}(j+\alpha_i)^{k}\bm {Z}'_{jk},\label{eq:cross_aligment_enc_B}
\end{align} where 
$\bm Z_{jk}$, $\bm Z^{'}_{jk}$ represent i.i.d., uniformly distributed noise terms to ensure privacy of $\bm A$ and $\bm B$. The scalars $\alpha_i,\forall i\in [1:N]$, are distinct elements of 
\begin{align}
	\label{eq_G}
	\mathbb{G}=\left\{\alpha_i\in \mathbb{F}\::\:j+\alpha_i\neq 0, \forall j\in [1:r]\right\}.
\end{align} 
The user then sends the pairs 
\begin{align*}
	(\tilde{\bm A}_{i}^{(1)},\tilde{\bm B}_{i}^{(1)}),\hdots,(\tilde{\bm A}_{i}^{(r)},\tilde{\bm B}_{i}^{(r)}) 
\end{align*} to the $i$-th server. The encoding in \eqref{eq:cross_aligment_enc_A} and \eqref{eq:cross_aligment_enc_B} is designed such that the decoding boils down to the interpolation of the (matrix) rational function
\begin{align}
	\bm F\left(\alpha\right)=\sum_{j=1}^{r}\frac{1}{\alpha-t_j}\bm X_j+\sum_{s=1}^{\bar{r}}\alpha^{j-1}\bm X_{r+s},\label{eq:mat_vec_rational_function}
\end{align} satisfying the conditions 
\begin{align}
	\bm F\left(\alpha_i\right)=\sum_{j=1}^{r}\tilde{\bm A}_{i}^{(j)}\tilde{\bm B}_{i}^{(j)},\quad i\in[1:N]
\end{align} for $r+\bar{r}\leq N$. Specifically, \eqref{eq:cross_aligment_enc_A} and \eqref{eq:cross_aligment_enc_B} ensure that in \eqref{eq:mat_vec_rational_function} 
\begin{align}
	\bm X_j&=\bm A\bm B_{j},\\
	\bm X_{r+s}&=\sum_{j=1}^{r}\sum_{\substack{k:k\geq s}}c_{j,k,s}\left(\bm A\bm Z'_{jk}+\bm Z_{jk}\bm B_{j}\right)+\sum_{j=1}^{r}\sum_{\substack{k,\tilde{k}:\\k+\tilde{k}\geq s}}d_{j,k+\tilde{k},s}\bm Z_{j\tilde{k}}\bm Z'_{jk}\label{eq:int_SCSA_dispersion}
\end{align} for $j\in[1:r]$ and $s\in[1:\bar{r}]$\footnote{The exact definition of the coefficients $c_{j,k,s}$ and $d_{j,k+\tilde{k},s}$ is not of importance and is thus omitted.}. 
In other words, through the encoding, we separate desired matrix products $\bm A\bm B_j$ from undesired matrix products (e.g., $\bm A\bm Z'_{jk}$) by assigning them, respectively, the rational or the polynomial part of $\bm F\left(\alpha\right)$. Interestingly, as opposed to other SDMM schemes in the literature, undesired matrix products are not solely attributed to a single power $\alpha^{j-1}$ but rather \emph{dispersed} to multiple powers.      

\subsection{Matrix Multiplication at the Servers}

Upon receiving all pairs $\left\{\tilde{\bm A}_{i}^{(j)},\tilde{\bm B}_{i}^{(j)}\right\}_{j=1}^{r}$, the $i$-th server computes
\begin{align}
	\bm O_{i}&=\sum_{j=1}^{r}\tilde{\bm A}^{(j)}_{i}\tilde{\bm B}_{i}^{(j)}\nonumber\\
		&=\begin{bmatrix}
			\frac{1}{1+\alpha_i} & \ldots &\frac{1}{r+\alpha_i} 
		\end{bmatrix}\otimes\bm I_{m}
		\begin{bmatrix}
			\left(\bm A\bm B_{1}\right)^{T} & \hdots & \left(\bm A\bm B_{r}\right)^{T}
		\end{bmatrix}^{T}\nonumber\\
	&\qquad+\begin{bmatrix}
		1 & \alpha_{i} & \ldots & \alpha_i^{2\ell-1}
	\end{bmatrix}\otimes\bm I_{m}
	\begin{bmatrix}
		\sum_{j=1}^{r}\bm{I}_{j,0}^{T} & \sum_{j=1}^{r}\bm{I}_{j,1}^{T} & \hdots & \sum_{j=1}^{r}\bm{I}_{j,2\ell-1}^{T}
	\end{bmatrix}^{T},\label{eq:answer_cross_subspace_scheme}
\end{align} 
where $\bm I_{jk}$\footnote{Note that $\bm X_{r+k+1}=\sum_{j=1}^{r}\bm I_{jk}$.} denotes the effective interference terms. The $i$-th server output $\bm O_i\in\mathbb{F}^{n\times\frac{p}{r}}$ is then transferred to the user. 

\subsection{Decoding at the User}

The user receives the server responses $\bm O_1,\ldots, \bm O_N$. In SCSA, \emph{all} undesired terms (e.g., $\bm A\bm Z_{jk}'$) disperse to multiple powers $\alpha_i^{k}$ in the range $k\in[0:2\ell-1]$. (The superposition of these terms gives $\bm I_{jk}$.) Simultaneously, all desired terms $\bm A\bm B_j$ are distinguishable from each other and from the interference by their unique powers $\frac{1}{j+\alpha_i}$. In other words, the user is able to decode the desired items since it can construct a \emph{full rank} 
decoding matrix 
\begin{align*}
	\bm D^{\text{SCSA}}=\begin{bmatrix}
		\frac{1}{1+\alpha_1} & \hdots & 		\frac{1}{r+\alpha_1} & 1 & \alpha_1 & \hdots & \alpha_1^{2\ell-1} \\
		\frac{1}{1+\alpha_2} & \hdots & \frac{1}{r+\alpha_2} & 1 & \alpha_2 & \hdots & \alpha_2^{2\ell-1}	\\
		\vdots & \vdots & \vdots & \vdots & \vdots & \vdots & \vdots \\
		\frac{1}{1+\alpha_N} & \hdots & \frac{1}{r+\alpha_N} & 1 & \alpha_N & \hdots & \alpha_N^{2\ell-1}	
	\end{bmatrix}
\end{align*} from server observations $\bm O_1,\ldots, \bm O_N$ each of dimension $n\times\nicefrac{p}{r}$ with $r=N-2\ell$. Thus, 
\begin{align*}
	K_{\text{DL}}^{\text{SCSA}}=\frac{N}{r}=\frac{N}{N-2\ell}.
\end{align*} Since we can reverse the partitioning in SCSA, we can show that
\begin{align*}
	K_{\text{UL}}^{\text{SCSA}(b)}=
		\begin{cases}
			K_{\text{UL}}^{\text{SCSA}(0)}=\frac{N\left(\frac{m}{p}+r\right)}{1+\frac{m}{p}}\quad&\text{ for }\mathsf{PART}\left([r,1],[1,1]\right)\\		
			K_{\text{UL}}^{\text{SCSA}(1)}=\frac{N\left(1+\frac{m}{p}r\right)}{1+\frac{m}{p}}\quad&\text{ for }\mathsf{PART}\left([1,1],[1,r]\right)
		\end{cases}.
\end{align*} Thus, the lowest possible uplink costs is the minimum of these two cases. Thus,
\begin{align*}
	K_{\text{UL}}^{\text{SCSA}}=\min_{b\in\{0,1\}}K_{\text{UL}}^{\text{SCSA}(b)}=\frac{N\min\left(1+\frac{m}{p}r,\frac{m}{p}+r\right)}{1+\frac{m}{p}}.
\end{align*} Note that the pairs $\left(K_{\text{UL}}^{\text{SCSA}(0)},K_{\text{DL}}^{\text{SCSA}}\right)$ and $\left(K_{\text{UL}}^{\text{SCSA}(1)},K_{\text{DL}}^{\text{SCSA}}\right)$ are both achievable in the UL-DL tradeoff curve.

\section{Uplink Cost Adjustable Schemes}
\label{sec:ul_ras}

In $\text{SCSA}(b)$, we can apply two partitioning/encoding scenarios -- $b=0$ and $b=1$. Recall that $\text{SCSA}(1)$ uses $\mathsf{PART}\left([1,1],[1,r]\right)$, i.e., the matrix $\bm A$ is left without partitioning while matrix $\bm B$ is horizontally partitioned into $r$ sub-matrices. The user conveys to the $i$-th server $r$ encoded pairs $\left(\tilde{\bm A}_i^{(j)},\tilde{\bm B}_{i}^{(j)}\right)$, $j\in[1:r]$. The transmission of multiple pairs with a low partitioning level to single servers results in an excessive use of uplink resources. To make better use of uplink resources, we propose two uplink cost adjustable SCSA schemes -- uplink-adjustable SCSA (USCSA) and group-based, uplink-adjustable SCSA (GSCSA) -- which guarantee better uplink performance than SCSA in exchange for an increase in the downlink costs. As opposed to the classical SCSA of section \ref{cha:rev_SCSA}, both schemes use a more general partitioning and a modified encoding strategy. We discuss the details in the next sub-sections. 

\subsection{Matrix Partitioning}

Table \ref{tab:mat_part} specifies the options of partitioning strategies applied by the user for the different schemes. We use the flag variable $b\in\{0,1\}$ to differentiate between $\mathsf{PART}\left([d,1],[1,c]\right)$ $(b=0)$ and $\mathsf{PART}\left([c,1],[1,d]\right)$ $(b=1)$. 
In $\text{GSCSA}(f,q,g,0)$, we assign sub-matrices $\bm A_j$, $j\in[1:fq]$ into $\bar{g}$ groups comprised of $g$ sub-matrices per group. Group $k,\forall k\in[1:\bar{g}]$, includes sub-matrices $\bm A_{[(k-1)g+1:kg]}\triangleq\left\{\bm A_{(k-1)g+1},\ldots,\bm A_{kg}\right\}$. In the sequel, we use the indexing set $\mathcal{I}_{k,g}\triangleq\{(k-1)g+1,\ldots,kg\}$ to refer to the $k$-th partitioning group comprised of $g$ elements. Throughout the text of this section, we use $g\in\left\{f,q\right\}$ and $\bar{g}=\left\{f,q\right\}\setminus g$.    

\begin{table}
	\caption{\footnotesize{Applied matrix partitioning strategies based on the proposed schemes. For all three schemes only one partitioning scenario is described in detail in the text of Section \ref{sec:ul_ras}.}}
	\label{tab:mat_part}
	\begin{center}
		\begin{tabular}{|l|l|l|l|}
			\hline
			Scheme & Sub-Scheme & Partitioning & Comments \\ \hline
				\multirow{2}{*}{$\text{SCSA}$} & $\text{SCSA}(0)$ & $\mathsf{PART}\left([r, 1],[1,1]\right)$ & \multirow{2}{*}{$r=N-2\ell$} \\ \cline{2-3} 
			& $\text{SCSA}(1)$ & $\mathsf{PART}\left([1, 1],[1,r]\right)$ & \\ \hline
				\multirow{2}{*}{$\text{USCSA}(f,q,g)$} & $\text{USCSA}(f,q,g,0)$ & $\mathsf{PART}\left(\left[g, 1\right],\left[1,\bar{g}\right]\right)$ & \multirow{2}{*}{$g\in\left\{f,q\right\}$, $\bar{g}=\left\{f,q\right\}\setminus g$} \\ \cline{2-3}
			& $\text{USCSA}(f,q,g,1)$ & $\mathsf{PART}\left(\left[\bar{g}, 1\right],\left[1,g\right]\right)$ & \\ \hline
				\multirow{2}{*}{$\text{GSCSA}(f,q,g)$} & $\text{GSCSA}(f,q,g,0)$ & $\mathsf{PART}\left([fq,1],[1,1]\right)$ & \multirow{2}{*}{$g\in\left\{f,q\right\}$, $\bar{g}=\left\{f,q\right\}\setminus g$} \\ \cline{2-3}
			& $\text{GSCSA}(f,q,g,1)$ & $\mathsf{PART}\left([1,1],[1,fq]\right)$ & \\ \hline 
		\end{tabular}
	\end{center}
\end{table}

\subsection{Encoding at the User}

Now, we describe the encoding of both $\text{USCSA}(f,q,g,0)$ and $\text{GSCSA}(f,q,v,0)$. Under the described partitioning, the encoded matrices destined to the $i$-th server are
\begin{itemize}[leftmargin=*]
	\item in case of $\text{USCSA}(f,q,g,0)$ for $j\in[1:\bar{g}]$
\end{itemize}
\begin{align}
	\tilde{\bm A}_i^{(j)} = &\sum\limits_{k=1}^{g}\frac{1}{k+(j-1)g+\alpha_i} \bm A_k+\sum\limits_{p=1}^{\ell} (j+\alpha_i)^{p-1} \bm Z_{jp},\label{eq:uscsa_enc_A} 
	\\
		\tilde{\bm B}_{i}^{(j)}= & \bm B_j+ \prod_{k=1}^{g} \big(k+(j-1)g+\alpha_i\big)\bigg(\sum\limits_{p=1}^{\ell} (j+\alpha_i)^{p-1} \bm Z'_{jp}\bigg), \label{eq:uscsa_enc_B}
\end{align} 
\begin{itemize}[leftmargin=*]
	\item and in case of $\text{GSCSA}(f,q,g,0)$ for $j\in[1:\bar{g}]$
\end{itemize}
\begin{align}
	\tilde{\bm A}_i^{(j)} = &\sum\limits_{k\in \mathcal{I}_{j,g}} \frac{1}{k+\alpha_i} \bm A_k+\sum\limits_{p=1}^{\ell} (j+\alpha_i)^{p-1} \bm Z_{jp},\label{eq:gscsa_enc_A}\\
		\tilde{\bm B}_i^{(j)} = & \bm B+ \prod_{k\in \mathcal{I}_{j,g}} (k+\alpha_i)\sum\limits_{p=1}^{\ell} (j+\alpha_i)^{p-1} \bm Z^{'}_{jp},\label{eq:gscsa_enc_B}
\end{align} where 
$\bm Z_{jp}$, $\bm Z'_{jp}$ represent i.i.d. noise terms to ensure privacy. The user then sends the pairs $\left(\tilde{\bm A}_i^{(j)},\tilde{\bm B}_{i}^{(j)}\right)_{j=1}^{\bar{g}}$ to the $i$-th server. 

\begin{remark}
	On the one hand, similarly to SCSA, the encoding in USCSA and GSCSA is designed to facilitate the dispersion of interference components to multiple subspaces which results in low download costs. On the other hand, simultaneously, both USCSA and GSCSA partition matrices $\bm A$ and $\bm B$ and accumulate them more efficiently than SCSA to encoding matrices $\tilde{\bm A}_i^{(j)}$ and $\tilde{\bm B}_i^{(j)}$, respectively. Overall, in comparison to SCSA, this results in lower upload costs at the expense of increased download costs. 
\end{remark}

\begin{remark}[USCSA vs. GSCSA]
	Recall that $\text{USCSA}(f,q,g)$ partitions both $\bm A$ and $\bm B$ into $f$ and $q$ parts, whereas $\text{GSCSA}(f,q,g)$ partitions either $\bm A$ or $\bm B$ into $fq$ parts. Since $fq\geq\max\left(f,q\right)$ for $f,g\in\mathbb{N}$, one can argue that with respect to upload costs USCSA (GSCSA) is better suited for matrices $\bm A\in\mathbb{F}^{m\times n},\bm F^{n\times p}$ where $m\approx p$ ($m\gg p$ or $m\ll p$) (cf. details in Section \ref{sec:comp_SDMM_schemes}).      
\end{remark}

\begin{remark}\label{rem:exchange_enc_roles}
	The encoding strategy of matrices $\bm A$ and $\bm B$ for both USCSA and GSCSA are exchangeable. In other words, we can use the encoding strategy in \eqref{eq:uscsa_enc_A} and \eqref{eq:gscsa_enc_A} for matrix $\bm B$ and the encoding in \eqref{eq:uscsa_enc_B} and \eqref{eq:gscsa_enc_B} for matrix $\bm A$. This establishes $\text{USCSA}(f,q,g,1)$ and $\text{GSCSA}(f,q,g,1)$. However, for the sake of brevity, we describe the remaining steps of the scheme for Eqs. \eqref{eq:uscsa_enc_A}-\eqref{eq:gscsa_enc_B}.    
\end{remark}

\subsubsection*{Privacy Constraint}

Now consider the privacy constraint for any $\ell$ colluding servers, given by $\left\{i_1,\ldots,i_\ell\right\}$ and $\bm \alpha_{\mathcal{L}}=[\alpha_{i_1},\ldots,\alpha_{i_\ell}]^{T}$ 
for both $\text{USCSA}(f,q,g,0)$ and $\text{GSCSA}(f,q,g,0)$. Further, we define $\bm Z\triangleq\left\{\bm Z_{j1},\ldots,\bm Z_{j\ell}\right\}_{j=1}^{\bar{g}}$ and $\bm Z'\triangleq$  $\left\{\bm Z'_{j1},\ldots,\bm Z'_{j\ell}\right\}_{j=1}^{\bar{g}}$\footnote{Note that these matrices differ in their dimension for USCSA and GSCSA. While for $\text{USCSA}(f,q,g,0)$ $[\bm Z]=\ell \bar{g}\frac{m}{g}\times n$, $[\bm Z']=\ell \bar{g} n\times\frac{p}{\bar{g}}$, for GSCSA the dimensions are $[\bm Z]=\ell\bar{g}\frac{m}{fq}\times n$, $[\bm Z']=\ell\bar{g}n\times p$.} for both USCSA and GSCSA. 
\begin{itemize}
	\item For $\text{USCSA}(f,q,g,0)$, 
	$\tilde{\bm A}_{\mathcal{L}}\triangleq\left\{\tilde{\bm A}_{i_1}^{(j)},\ldots,\tilde{\bm A}_{i_\ell}^{(j)}\right\}_{j=1}^{\bar{g}}$ and $\tilde{\bm B}_{\mathcal{L}}\triangleq\left\{\tilde{\bm B}_{i_1}^{(j)},\ldots,\tilde{\bm B}_{i_\ell }^{(j)}\right\}_{j=1}^{\bar{g}}$ can be represented by
\end{itemize}	
\begin{align}
\tilde{\bm A}_{\mathcal{L}}&=\bm I_{\bar{g}}\otimes\bm I_{\ell}\cdot\left(\mathsf{blkC}\left(\bm \alpha_{\mathcal{L}},g,\bar{g}\right)\otimes\bm I_{\frac{m}{g}}\cdot				
\underbrace{\begin{bmatrix}
	\bm A_1 \\ \bm A_2 \\ \vdots \\ \bm A_g
	\end{bmatrix}}_{=\bm A}
+\mathsf{diagV}\left(\bm \alpha_{\mathcal{L}},\bar{g}\right)\otimes\bm I_{\frac{m}{g}}\cdot\bm Z\right),\label{eq:col_server_A_USCSA}\\
\tilde{\bm B}_{\mathcal{L}}&=\bm I_{\bar{g}}\otimes \bm 1_{\ell}\otimes \bm I_{n}\cdot
\underbrace{\begin{bmatrix}
	\bm B_1 \\ \bm B_2 \\ \vdots \\ \bm B_{\bar{g}}
	\end{bmatrix}}_{\triangleq \bm B'}+\bm \Gamma(\bar{g})\cdot\mathsf{diagV}\left(\bm \alpha_{\mathcal{L}},\bar{g}\right)\otimes \bm I_{n}\cdot\bm Z',\label{eq:col_server_B_USCSA}
\end{align} where
\begin{align*}
	\mathsf{blkC}\left(\bm \alpha_{\mathcal{L}},g,\bar{g}\right)&=
	\begin{bmatrix}
	\mathsf{C}\left(\bm \alpha_{\mathcal{L}},-[1:g]\right) \\
	\mathsf{C}\left(\bm \alpha_{\mathcal{L}},-[g+1:2g]\right) \\
	\vdots \\
	\mathsf{C}\left(\bm \alpha_{\mathcal{L}},-[(\bar{g}-1)g+1:\bar{g}g]\right)		
	\end{bmatrix},\\
	\mathsf{diagV}\left(\bm \alpha_{\mathcal{L}},\bar{g}\right)&=
	\begin{bmatrix}
	\mathsf{V}_\ell\left(\bm 1+\bm \alpha_{\mathcal{L}}\right) & \bm 0_{\ell} & \hdots & \bm 0_{\ell} \\
	\bm 0_{\ell} & \mathsf{V}_\ell\left(\bm 2+\bm \alpha_{\mathcal{L}}\right) & \hdots & \bm 0_{\ell} \\ \Large{\vdots} & \Large{\vdots} & \Large{\ddots} & \Large{\vdots} \\
	\bm 0_{\ell} & \bm 0_{\ell} & \hdots & \mathsf{V}_\ell\left(\bar{\bm{g}}+\bm \alpha_{\mathcal{L}}\right)
	\end{bmatrix},
\end{align*}
\begin{align*}
\bm\Gamma(\bar{g})&=
	\begin{bmatrix}
		\prod_{j=1}^{g}\mathsf{diag}\left(\bm j+\bm \alpha_{\mathcal{L}}\right) & \hdots & \bm 0_{\ell} \\
		\Large{\vdots} & \Large{\ddots} & \Large{\vdots} \\
		\bm 0_{\ell} & \hdots & \prod_{j=1}^{g}\mathsf{diag}\left(\bm j+(\bar{g}-1)\bm g+\bm \alpha_{\mathcal{L}}\right)
	\end{bmatrix},
\end{align*} with the Cauchy and Vandermonde matrices being
\begin{align*}
	\mathsf{C}(\bm a, \bm b)&=
	\begin{bmatrix}
		\frac{1}{a_1-b_1} & \frac{1}{a_1-b_2} & \frac{1}{a_1-b_3} & \hdots & \frac{1}{a_1-b_n} \\
		\frac{1}{a_2-b_1} & \frac{1}{a_2-b_2} & \frac{1}{a_2-b_3} & \hdots & \frac{1}{a_2-b_n} \\
		\vdots & \vdots & \vdots & \ddots &	\vdots \\
		\frac{1}{a_m-b_1} & \frac{1}{a_m-b_2} & \frac{1}{a_m-b_3} & \hdots & \frac{1}{a_m-b_n} \\							
	\end{bmatrix}\in\mathbb{F}^{m\times n}\\
	\mathsf{V}_{n}(\bm a)&=
	\begin{bmatrix}
		1 & a_1 & a_1^{2} & \hdots & a_1^{n-1} \\
		1 & a_2 & a_2^{2} & \hdots & a_2^{n-1} \\
		\vdots & \vdots & \hdots & \ddots &	\vdots \\
		1 & a_m & a_m^{2} & \hdots & a_m^{n-1} \\							
	\end{bmatrix}\in\mathbb{F}^{m\times n} 
\end{align*} for $\bm a=[a_1,a_2,\ldots,a_m]^{T}$ and $\bm b=[b_1,b_2,\ldots,b_n]^{T}$.
\begin{itemize}		
	\item Similarly, for GSCSA we can write the observations $\tilde{\bm A}_{\mathcal{L}}\triangleq\left\{\tilde{\bm A}_{i_1}^{(j)},\ldots,\tilde{\bm A}_{i_\ell}^{(j)}\right\}_{j=1}^{\bar{g}}$ and $\tilde{\bm B}_{\mathcal{L}}\triangleq\left\{\tilde{\bm B}_{i_1}^{(j)},\ldots,\tilde{\bm B}_{i_\ell}^{(j)}\right\}_{j=1}^{\bar{g}}$ as follows
\end{itemize}	
\begin{align}
	\tilde{\bm A}_{\mathcal{L}}&=\bm I_{\bar{g}}\otimes\bm I_{\ell}\cdot\left(\mathsf{diagC}\left(\bm \alpha_{\mathcal{L}},g,\bar{g}\right)\otimes\bm I_{\frac{m}{fq}}\cdot				
	\underbrace{\begin{bmatrix}
		\bm A_1 \\ \bm A_2 \\ \vdots \\ \bm A_{fq}
		\end{bmatrix}}_{=\bm A}
	+\mathsf{diagV}\left(\bm \alpha_{\mathcal{L}},\bar{g}\right)\otimes\bm I_{\frac{m}{fq}}\cdot\bm Z\right),\label{eq:col_server_A_GSCSA} \\
		\tilde{\bm B}_{\mathcal{L}}&=\bm 1_{\bar{g}\ell}\otimes \bm B + \bm \Gamma(\bar{g})\cdot\mathsf{diagV}\left(\bm \alpha_{\mathcal{L}},\bar{g}\right)\otimes \bm I_{n}\cdot\bm Z',\label{eq:col_server_B_GSCSA}
\end{align} where
\begin{align*}
\mathsf{diagC}&\left(\bm \alpha_{\mathcal{L}},g,\bar{g}\right)=\\
	&\begin{bmatrix}
		\mathsf{C}\left(\bm \alpha_{\mathcal{L}},-[1:g]\right) & \bm 0_{\ell} & \hdots & \bm 0_{\ell} \\
		\bm 0_{\ell} & \mathsf{C}\left(\bm \alpha_{\mathcal{L}},-[g+1:2g]\right) & \hdots & \bm 0_{\ell} \\ 
		\Large{\vdots} & \Large{\vdots} & \Large{\ddots} & \Large{\vdots} \\
		\bm 0_{\ell} & \bm 0_{\ell} & \hdots & \mathsf{C}\left(\bm \alpha_{\mathcal{L}},-[(\bar{g}-1)v+1:\bar{g}g]\right)	
	\end{bmatrix}.
\end{align*} From \eqref{eq:col_server_A_USCSA}-\eqref{eq:col_server_B_GSCSA}, it is possible to rewrite $\tilde{\bm A}_{\mathcal{L}}$ and $\tilde{\bm B}_{\mathcal{L}}$ compactly by 
\begin{align}
	\tilde{\bm A}_{\mathcal{L}}&=\bm P\cdot\left(\bm Q\otimes \bm I\cdot \bm A+\bm R\otimes\bm I\cdot\bm Z\right),\label{eq:col_server_A}\\
	\tilde{\bm B}_{\mathcal{L}}&=\bm S\left(\bm B\right)+\bm T\otimes\bm I\cdot \bm Z',\label{eq:col_server_B}
\end{align} where
\begin{align*}
	\bm S\left(\bm B\right)=
	\begin{cases}
	\bm I_{q}\otimes \bm 1_{\ell}\otimes \bm I_{n}\cdot\bm B'&\text{ for USCSA} \\
	\bm 1_{q\ell}\otimes \bm B&\text{ for GSCSA} 
	\end{cases}.
\end{align*} Since the inverses for $\bm P$ and $\bm R\otimes \bm I$ in \eqref{eq:col_server_A} and $\bm T\otimes\bm I$ in \eqref{eq:col_server_B} exist, we can show that the observations of any $\ell$ colluding servers are independent of $\bm A$ and $\bm B$.
\begin{align*}
	I&\left(\bm A,\bm B;\tilde{\bm A}_{\mathcal{L}},\tilde{\bm B}_{\mathcal{L}}\right)=I\left(\bm A,\bm B;\left(\bm R\otimes \bm I\right)^{-1}\bm P^{-1}\tilde{\bm A}_{\mathcal{L}},\left(\bm T\otimes\bm I\right)^{-1}\tilde{\bm B}_{\mathcal{L}}\right)\\
	&\stackrel{\eqref{eq:col_server_A},\eqref{eq:col_server_B}}{=}I\left(\bm A,\bm B;\left(\bm R\otimes \bm I\right)^{-1}\bm Q\otimes \bm I\cdot\bm A+\bm Z,\left(\bm T\otimes \bm I\right)^{-1}\bm S(\bm B)+\bm Z'\right)\leq I(\bm A,\bm B; \bm Z, \bm Z')=0
\end{align*} Since $I\left(\bm A,\bm B;\tilde{\bm A}_{\mathcal{L}},\tilde{\bm B}_{\mathcal{L}}\right)\geq 0$ in general, we infer that 
\begin{align*}
	I\left(\bm A,\bm B;\tilde{\bm A}_{\mathcal{L}},\tilde{\bm B}_{\mathcal{L}}\right)=0.
\end{align*}

\subsection{Matrix Multiplication at the Servers}

Every server $i$ multiplies its pairs and accumulates them to retrieve the server output $\bm O_i$. Mathematically, the server output becomes for both $\text{USCSA}(f,q,g,0)$ and $\text{GSCSA}(f,q,v,0)$ 
\begin{align*}
	\bm O_{i}&=\sum_{j=1}^{\bar{g}}\tilde{\bm A}^{(j)}_{i}\tilde{\bm B}_{i}^{(j)}=\sum_{j=1}^{\bar{g}}\bm C_{ji},
\end{align*} where e.g. for $\text{USCSA}(f,q,g,0)$ we may show 
\begin{align*}
	\bm C_{ji}&=\sum_{k=1}^{g}\frac{1}{k+(j-1)g+\alpha_i}\bm A_{k}\bm B_j+\sum_{t=0}^{2(\ell-1)+g}\alpha_i^{t}\bm I_{jt}.
\end{align*} Note that for the derivation, we use $(a)$ the binomial expansion in the form $(j+\alpha_i)^{p}=\sum_{t=0}^{p}{p\choose t}\alpha_i^{t}j^{p-t}$ and $(b)$ combine all undesired terms $\bm A_{k}\bm Z'_{jp}$, $\bm Z_{jp}\bm B_{j}$ and $\bm Z_{jp'}\bm Z'_{jp}$ with matching exponents in $\alpha_i$ to effective interference components $\bm I_{jt}$, $\forall t\in[0:2\left(\ell-1\right)+g]$. 
A similar expression on $\bm C_{ji}$ can be derived for GSCSA but is omitted here for the sake of brevity. Overall, summing $\bm C_{ji}$ over $j\in[\bar{g}]$, 
we obtain
\begin{itemize}
	\item for $\text{USCSA}(f,q,g,0)$
	\begin{align*}
		\bm O_{i}=\sum_{j=1}^{\bar{g}}\sum_{k=1}^{g}\frac{1}{k+(j-1)g+\alpha_i}\bm A_{k}\bm B_j+\sum_{j=1}^{\bar{g}}\sum_{k=0}^{2(\ell-1)+g}\alpha_i^{k}\bm I_{jk}
	\end{align*}
	\item and for $\text{USCSA}(f,q,g,0)$
	\begin{align*}
		\bm O_{i}= \sum\limits_{j=1}^{\bar{g}}\sum\limits_{k=1}^{g}\frac{1 }{k+(j-1)g+
		\alpha_i} \bm A_{(j-1)g+k} \bm B
		+ & \sum\limits_{j=1}^{\bar{g}}\sum\limits_{k=0}^{2(\ell-1)+g}\alpha_{i}^k \bm I_{jk}.
	\end{align*}
\end{itemize}

\subsection{Decoding at the User}

Similarly to SCSA, we can derive a decoding matrix for both USCSA and GSCSA by finding a linear representation of $\bm O_{[1:N]}$ as a function of desired sub-matrix products $\bm A_{k}\bm B_{j}$ and interfering terms $\bm I_{k}'=\sum_{j}\bm I_{jk}$. The general \emph{full-rank} decoding matrix for both USCSA and GSCSA is given by
\begin{align*}	
	\bm D=\begin{bmatrix}
			\frac{1}{(1+\alpha_1)} & \hdots & \frac{1}{(fq+\alpha_1)} & 1 & \alpha_1 & \hdots & \alpha_1^{2(\ell-1)+g} \\
			\frac{1}{(1+\alpha_2)} & \hdots & \frac{1}{(fq+\alpha_2)} & 1 & \alpha_2 & \hdots & \alpha_2^{2(\ell-1)+g}	\\
			\vdots & \vdots & \vdots & \vdots & \vdots & \vdots & \vdots \\
			\frac{1}{1+\alpha_N} & \hdots & \frac{1}{(fq+\alpha_N)} & 1 & \alpha_N & \hdots & \alpha_N^{2(\ell-1)+g}	
	\end{bmatrix} 
\end{align*} for $g\in\{f,q\}$. This matrix has the dimension $N\times Q$ with
\begin{align*}
Q^{\text{USCSA}(f,q,g,b)}=Q^{\text{GSCSA}(f,q,g,b)}&=fq+g+2\ell-1
\end{align*} for $b\in\{0,1\}$ and $g\in\{f,q\}$\footnote{Since $Q$ is independent of $b\in\{0,1\}$, in the sequel, we remove the index $b$ wherever possible.}. Since there are at most $N$ server observations accessible, the user selects $f$, $q$, and $g$ such that $Q\leq N$. Overall the user recovers $fq$ desired items from $Q$ overall items (including $2\ell+g-1$, $g\in\{f,q\}$, interference terms $\bm I'_{k}$). Thus, we attain the download costs  
\begin{align}
	K_{\text{DL}}^{\text{USCSA}(f,q,g)}&=\frac{fq+g+2\ell-1}{fq},\label{eq:KDL_USCSA}\\
	K_{\text{DL}}^{\text{GSCSA}(f,q,g)}&=\frac{fq+g+2\ell-1}{fq}.\label{eq:KDL_GSCSA}   
\end{align} For these downlink costs we need the following respective uplink costs.
\begin{align}
	K_{\text{UL}}^{\text{USCSA}(f,q,g)}=\min_{b\in\{0,1\}}K_{\text{UL}}^{\text{USCSA}(f,q,g,b)}&=\min\left(\frac{N\left(1+\frac{m}{p}\frac{\bar{g}}{g}\right)}{1+\frac{m}{p}},\frac{N\left(\frac{\bar{g}}{g}+\frac{m}{p}\right)}{1+\frac{m}{p}}\right),\label{eq:KUL_USCSA}\\
		K_{\text{UL}}^{\text{GSCSA}(f,q,g)}=\min_{b\in\{0,1\}}K_{\text{UL}}^{\text{GSCSA}(f,q,g,b)}&=\min\left(\frac{\frac{N}{f}\left(\bar{g}f+\frac{m}{p}\frac{\bar{g}}{q}\right)}{1+\frac{m}{p}},\frac{\frac{N}{f}\left(\frac{\bar{g}}{q}+\frac{m}{p}\bar{g}f\right)}{1+\frac{m}{p}}\right).\label{eq:KUL_GSCSA}	
\end{align} 
Ultimately, we can flexibly balance the matrix partitioning (and ultimately the uplink costs) against the downlink costs. For the special case, when $g=1$ and $\bar{g}=fq=N-2\ell$, USCSAs downlink and uplink costs reduce to the one of SCSA.

\section{Comparison of SDMM Schemes}  
\label{sec:comp_SDMM_schemes}

In the following, we first construct a lower bound on $K_{\text{UL}}$. 
\begin{lemma}\label{lem:conv_K_UL}
	For independent and uniformly distributed matrices $\bm A$ and $\bm B$, the uplink cost $K_{\text{UL}}$ is bounded from below by
	\begin{align}
	K_{\text{UL}}\geq\frac{N}{N-\ell}.\label{eq:bound_K_UL}
	\end{align}
\end{lemma}
\begin{proof}
	Details of the proof are provided in the Appendix. 
\end{proof}

\subsection{USCSA vs. GSCSA}

The uplink costs for both USCSA and GSCSA are  $\nicefrac{m}{p}$ for $\nicefrac{m}{p}\in(0,1]$ and decreasing in $\nicefrac{m}{p}$ for $\nicefrac{m}{p}\in[1,\infty)$.
Thus, defining $\underline{K}_{\text{UL}}\triangleq\inf_{\nicefrac{m}{p}} K_{\text{UL}}$ and $\overline{K}_{\text{UL}}\triangleq\sup_{\nicefrac{m}{p}} K_{\text{UL}}$,  
we get for USCSA $\underline{K}_{\text{UL}}^{\text{USCSA}(f,q,g)}=N\min\left(1,\frac{\bar{g}}{g}\right)$, $\overline{K}_{\text{UL}}^{\text{USCSA}(f,q,g)}=\frac{N\left(1+\frac{\bar{g}}{g}\right)}{2}$
and for GSCSA $\underline{K}_{\text{UL}}^{\text{GSCSA}(f,q,g)}=N\frac{\bar{g}}{fq}$, $	\overline{K}_{\text{UL}}^{\text{GSCSA}(f,q,g)}=\frac{N\bar{g}\left(1+\frac{1}{fq}\right)}{2}$.
Recall that $K_{\text{DL}}^{\text{USCSA}(f,q,g)}=K_{\text{DL}}^{\text{GSCSA}(f,q,g)}$ while $K_{\text{UL}}^{\text{USCSA}(f,q,g)}\neq K_{\text{UL}}^{\text{GSCSA}(f,q,g)}$ in general for $g\neq 1$ or $\bar{g}\neq 1$. 
If $g,\bar{g}>1$, we infer from \eqref{eq:KUL_USCSA} and \eqref{eq:KUL_GSCSA} that for
\begin{itemize}
	\item $\nicefrac{m}{p}\in\left(0,\nicefrac{1}{\max\left(f,q\right)}\right]\cup\left(\max\left(f,q\right),\infty\right)$: $K_{\text{UL}}^{\text{GSCSA}(f,q,g)}\leq K_{\text{UL}}^{\text{USCSA}(f,q,g)}$ and
	\item $\nicefrac{m}{p}\in\left(\nicefrac{1}{\max\left(f,q\right)},\max\left(f,q\right)\right]$: $K_{\text{UL}}^{\text{GSCSA}(f,q,g)}\geq K_{\text{UL}}^{\text{USCSA}(f,q,g)}$.  
\end{itemize} In other words, for low and high ratios of $\nicefrac{m}{p}$, $\text{GSCSA}(f,q,g)$ outperforms $\text{USCSA}(f,q,g)$. Further, we may upper bound the additive gap
\begin{align}
	0\leq\frac{1}{K^{\star}_{\text{UL}}}-\frac{1}{K_{\text{UL}}}&\leq\frac{N-\ell}{N}-\frac{1}{\overline{K}_{\text{UL}}}
	=1-\frac{\ell}{N}-
	\begin{cases}
	\frac{2}{N\left(1+\frac{\bar{g}}{g}\right)}\quad&\text{for USCSA}(f,q,g)\\
		\frac{2}{N\left(\bar{g}+\frac{1}{g}\right)}\quad&\text{for GSCSA}(f,q,g)
	\end{cases}\nonumber\\
	&\leq 1-\frac{\ell}{N}-
	\begin{cases}
	\frac{2}{N\left(1+\frac{\max\left(f,q\right)}{\min\left(f,q\right)}\right)}\quad&\text{for USCSA}(f,q,g)\\
	\frac{2}{N\left(\max\left(f,q\right)+\frac{1}{\min\left(f,q\right)}\right)}\quad&\text{for GSCSA}(f,q,g)
	\end{cases}\nonumber\\		
		&\stackrel{(a)}{\leq} 1-\frac{\ell}{N}-\frac{2}{N\left(N-2\ell+1\right)}=1-\frac{\ell}{N}-\frac{2}{N^{2}}\cdot\frac{1}{1-\frac{(2\ell-1)}{N}},		
\end{align} where $(a)$ follows from maximizing the denominator of the last term for $Q^{\text{USCSA}},Q^{\text{GSCSA}}\leq N$. This suggests that the maximum additive gap $\frac{1}{K^{\star}_{\text{UL}}}-\frac{1}{K_{\text{UL}}}$ is decreasing in $\nicefrac{\ell}{N}$.

\subsection{Uplink-Downlink-Cost-Tradeoff -- Comparison with Other Schemes}

\begin{figure}[h]
	\centering
	\input{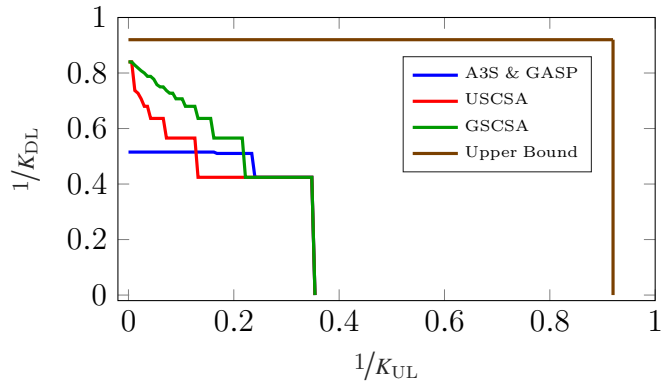}
	\caption{\footnotesize Comparison between the achievable reciprocal uplink and downlink cost pairs for multiple schemes: (i) A3S \& GASP Codes, (ii) USCSA and (iii) GSCSA. In these plots, we set $N=100$, $\ell=8$ and $\nicefrac{m}{p}=200$. For the upper bounds, we use $\nicefrac{1}{K_{\text{DL}}}\leq\nicefrac{(N-\ell)}{N}=0.92$ and $\nicefrac{1}{K_{\text{UL}}}\leq\nicefrac{(N-\ell)}{N}=0.92$.}
	\label{plot:Rate_over_ell}
\end{figure}

We now compare the achievable reciprocal uplink and downlink cost pairs $\left(\nicefrac{1}{K_{\text{UL}}},\nicefrac{1}{K_{\text{DL}}}\right)$ for various SDMM schemes. These are the aligned secret sharing scheme (A3S) \cite{Kakar_2018} and the gap additive secure polynomial (GASP) scheme \cite{D'Oliveira_2018}. Fig. \ref{plot:Rate_over_ell} shows the performance of all schemes when $N=100$, $\ell=8$, $g=f$ and $\nicefrac{m}{p}=200$.          

Recall that in this case, both $\nicefrac{1}{K_{\text{UL}}}$ and $\nicefrac{1}{K_{\text{DL}}}$ are bounded from above by $\nicefrac{(N-\ell)}{N}=0.92$ (cf. dotted line in Fig. \ref{plot:Rate_over_ell}). On the one hand, the lowest downlink cost of $K_{\text{DL}}=\nicefrac{1}{0.84}$ is attained for SCSA (or USCSA and GSCSA when $f=1$ and $q=N-2\ell=84$) when $\nicefrac{1}{K^{\text{SCSA}}_{\text{UL}}}\leq\nicefrac{1}{141.3}\approx 0.007$. On the other hand, the lowest uplink cost is when $\frac{1}{K_{\text{UL}}}\approx 0.35$ for which $K_{\text{DL}}\leq\frac{42}{99}\approx 0.42$. For USCSA and GSCSA, this cost pair is achievable for $f=42$ and $q=1$. For this example, we see that for almost all pairs, GSCSA outperforms all remaining schemes.

\section{Design of Polynomial Evaluation Points}

So far, we have focused solely on the communication aspect of CSA schemes. In this section, we propose an efficient polynomial evaluation scheme suited for the CSA encoding. 

\subsection{Polynomial Evaluation at a Single Point}

Consider the evaluation of an $(n-1)$-th degree polynomial 
\begin{align}
	u(x)=u_{n-1} x^{n-1}+u_{n-2}x^{n-2}+\ldots+u_{1}x+u_{0},\quad u_{n-1}\neq 0\label{eq:single_evaluation_polynomial}
\end{align} at some point $x$. 
One may decompose \eqref{eq:single_evaluation_polynomial} in the form,
\begin{align}
	u(x)=u_{\text{even}}\left(x^{2}\right)+xu_{\text{odd}}\left(x^{2}\right),\label{eq:polynom_decomposition}
\end{align} where 
\begin{align*}
	u_{\text{even}}\left(x\right)&=\sum_{i=0}^{\floor{\nicefrac{(n-1)}{2}}}u_{2i}x^{i},\\
	u_{\text{odd}}\left(x\right)&=\sum_{i=0}^{\floor{\nicefrac{n}{2}}-1}u_{2i+1}x^{i},
\end{align*} are, respectively, the even and odd subsidiary polynomials. The procedure
\begin{align}
	u(x)=\left(\ldots\left(u_{n-1}x+u_{n-2}\right)x+\cdots\right)x+u_{0},
\end{align} that is starting with $u_{n-1}$ multiplying by $x$, adding $u_{n-2}$, multiplying by $x$, repeating this procedure until finally adding $u_{0}$ gives $u(x)$. In the literature, it is known by \emph{Horner's rule} (HR) which requires $n-1$ multiplications and $n-1$ additions, minus one addition for each coefficient that is zero. The Horner scheme is optimal in the sense that a polynomial of degree $n-1$ cannot be evaluated with fewer arithmetic operations \cite{Blaeser13}. In the sequel of this chapter, the 'unit' \emph{operation} means a multiplication followed by an addition. Note that in general the complexity of evaluating $p$ polynomials $u_m(x)$ of degree $\deg\left(u_m(x)\right)=n_i$ at a single point $x=\bar{x}$ according to HR is $C_{\text{EC}}^{\text{HR}}\left(n_1,\ldots,n_p\right)=\sum_{m=1}^{p}n_m$ operations. 

The strategy that separately applies Horner's rule on $u_{\text{even}}\left(x^{2}\right)$ and $xu_{\text{odd}}\left(x^{2}\right)$ and combines them according to \eqref{eq:polynom_decomposition} to $u(x)$ is known as the \emph{second-order} (SO) Horner's rule (SO-HR) \cite{Knuth97}. The SO-HR scheme uses $n$ multiplications and $n-1$ additions to determine $u(x)$\footnote{We count the evaluation of $x^{2}$ as one extra multiplication.}. This corresponds to approximately $n$ (scalar) operations.     

\subsection{Polynomial Evaluation at $n$ Points}   
\label{subsec:multi_point_poly_eval}

Now consider the same polynomial $u(x)$ that shall be evaluated at $n$ points $x_{0},x_{1},\ldots,x_{n-1}$. 
On closer inspection of \ref{eq:polynom_decomposition}, one may realize that the cost for evaluating $u(-x)$ from $u_{\text{even}}(x)$ and $xu_{\text{odd}}(x)$ is only one additional multiplication. Using the SO strategy in conjuction with HR for the evaluation of $n$ points, one needs in total $n\left(n-\floor{\nicefrac{n}{2}}\right)$ (instead of $n(n-1)$ for HR) operations.   

\subsection{Second-Order Encoding of CSA schemes}

For different matrix coefficients, $\bar{j}$\footnote{For SCSA $\bar{j}=r$, USCSA and GSCSA $\bar{j}\in\left\{f,q\right\}$.} matrix polynomials, each evaluated at $\alpha=\alpha_i$, i.e.,   
\begin{align}
	\bm U^{(j)}\left(\alpha_i\right)=\sum_{p=0}^{d-1}\bm U_{jp}\left(j+\alpha_i\right)^{p},
\end{align} $\forall i\in[1:N]$, $\forall j\in\left[1:\bar{j}\right]$ represent intermediate results of CSA encoding matrices $\tilde{\bm A}_{i}^{(j)}$ and $\tilde{\bm B}_{i}^{(j)}$. For these particular matrix polynomials, we choose $\bm \alpha=[\alpha_1,\ldots,\alpha_N]^{T}$ with $\alpha_i\in\mathbb{G}$ such that SO-based polynomial evaluation (cf. \ref{subsec:multi_point_poly_eval}) of $\bm U^{(j)}$ is applicable. Implicitly it is assumed that $\bar{j},N\geq 2$. Define $\bar{y}_D=\floor{\nicefrac{N}{(\bar{j}+1)}}$ and $N_R=N-\bar{y}_D(\bar{j}+1)$\footnote{Note that $N_R\leq\bar{j}$.}. For $\alpha_i\in\mathbb{G}$, $\forall i\in[1:N]$ and distinct $\left(\alpha_{s(\bar{j}+1)+1}\right)_{s=0}^{\bar{y}_D}$, the relationship 
\begin{align}
	\begin{cases}
		j+\alpha_{\left(s-1\right)\left(\bar{j}+1\right)+1}=-\left(j+\alpha_{\left(s-1\right)\left(\bar{j}+1\right)+j+1}\right),\quad&\forall j\in\left[1:\bar{j}\right],\forall s\in\left[1:\bar{y}_D\right]\\
			j+\alpha_{\bar{y}_D\left(\bar{j}+1\right)+1}=-\left(j+\alpha_{\bar{y}_D\left(\bar{j}+1\right)+j+1}\right),&\forall j\in\left[1:N_{R}\right]
	\end{cases}\label{eq:choice_alpha_SO}	
\end{align} generates in total $N_{R}+\bar{y}_D\bar{j}-1=N-\bar{y}_D-1$ additive inverse pairs $\left(x,-x\right)$ for which SO-HR can be used in the encoding process. In Table \ref{tab:enc_complexity_derivation}, we derive the overall encoding complexity $C_{\text{EC}}^{(\text{SO-HR CSA})}$ of evaluating the matrix polynomial $\bm U^{(j)}$, $\forall j\in\left[1:\bar{j}\right]$ at $N$ points $\alpha=\alpha_i$ chosen according to \eqref{eq:choice_alpha_SO}.  $|\bm U|$ denotes the number of matrix elements of $\bm U^{(j)}$, $\forall j\in\left[1:\bar{j}\right]$. 

\begin{table}[h]
	\caption{\small{Derivation of the encoding complexity of evaluating the matrix polynomial $\bm U^{(j)}$, $\forall j\in\left[1:\bar{j}\right]$ at $N$ points $\alpha=\alpha_i$, $\forall i\in[1:N]$.}}
	\begin{center}
	\footnotesize{
		\begin{tabular}{ | c | c | c| c | }
			\hline
			\multirow{2}{*}{Category} & \multirow{2}{*}{$\#$} & \multirow{2}{*}{\shortstack{Complexity \\ $\left[\text{operation}\right]$}} & \multirow{2}{*}{\shortstack{Overall \\ $\left[\text{operation}\right]$}} \\ 
			& & &  \\ \hline 
			SO-Pair & $N-\bar{y}_D-1$ & $\approx \left(d-1+\lambda_{+}\right)|\bm U|$\footnote{Typically, since $|\bm U|\gg 1$, the scalar multiplication of $x^{2}$ at evaluation point $x$ is ignored in the operation count.} & $\left(N-\bar{y}_D-1\right)\left(d-1+\lambda_{+}\right)|\bm U|$ \\ \hline
			No pair & $N\left(\bar{j}-2\right)+2\left(\bar{y}_D+1\right)$ & $\left(d-1\right)|\bm U|$ & $\left(N\left(\bar{j}-2\right)+2\left(\bar{y}_D+1\right)\right)\left(d-1\right)|\bm U|$ \\ \hline
			-- & -- & -- & $\left(N\bar{j}\left(d-1\right)-\left(d-1-\lambda_{+}\right)\left(N-\bar{y}_D-1\right)\right)|\bm U|$ \\ \hline
	\end{tabular}}
\end{center}
	\label{tab:enc_complexity_derivation}
\end{table}

The final expression is as follows
\begin{align}
	C_{\text{EC}}^{(\text{SO-HR CSA})}\left(|\bm U|,d,N,\bar{j}\right)=\left(N\bar{j}\left(d-1\right)-\left(d-1-\lambda_{+}\right)\left(N-\bar{y}_D-1\right)\right)|\bm U|.\label{eq:enc_complexity_polynom_u}
\end{align} In this equation $\lambda_{+}\in(0,1)$ (similarly $\lambda_{\text{\textbullet}}$) denotes the relative time needed for a scalar addition (multiplication) in comparison to a scalar operation\footnote{$\lambda_{+}+\lambda_{\text{\textbullet}}=1$}.

We can now derive the complexity of the CSA schemes. Consider first the encoding of $\text{SCSA}(1)$ (cf. \eqref{eq:cross_aligment_enc_A} and \eqref{eq:cross_aligment_enc_B}). The encoding of $\left(\tilde{\bm A}_{i}^{(j)}\right)_{i=1,j=1}^{i=N,j=r}$, on the one hand, can be realized through SO-HR (following \eqref{eq:choice_alpha_SO}) and a subsequent multiplication with $\frac{1}{j+\alpha_i}$\footnote{We ignore the effort of calculating $N\bar{j}$ multiplicative inverses of $j+\alpha_i$, $j\in[1:\bar{j}]$, $i\in[1:N]$, since typically $|\bm U|\gg N\bar{j}$.}. Overall, this requires $	C_{\text{EC}}^{(\text{SO-HR CSA})}\left(|\bm A|,\ell+1,N,r\right)+\lambda_{\text{\textbullet}}rN|\bm A|$
operations. The encoding of $\left(\tilde{\bm B}_{i}^{(j)}\right)_{i=1,j=1}^{i=N,j=r}$, on the other hand is attainable using SO-HR with $C_{\text{EC}}^{(\text{SO-HR CSA})}\left(\nicefrac{|\bm B|}{r},\ell+1,N,r\right)$
operations. Thus, the overall encoding complexity of $\text{SCSA}(1)$ using SO-HR becomes
\begin{align}
	C_{\text{EC}}^{\text{SO-HR SCSA}(1)}=	C_{\text{EC}}^{(\text{SO-HR CSA})}\left(|\bm A|+\nicefrac{|\bm B|}{r},\ell+1,N,r\right)+\lambda_{\text{\textbullet}}rN|\bm A|.\label{eq:enc_complexity_SCSA_1}
\end{align} Recall that in the encoding of $\text{SCSA}(0)$, $\tilde{\bm A}_{i}^{(j)}$ ($\tilde{\bm B}_{i}^{(j)}$) is encoded as $\tilde{\bm B}_{i}^{(j)}$ ($\tilde{\bm A}_{i}^{(j)}$) for $\text{SCSA}(1)$. This implies that the encding complexity of $\text{SCSA}(0)$ applying SO-HR is
\begin{align}
	C_{\text{EC}}^{\text{SO-HR SCSA}(0)}=	C_{\text{EC}}^{(\text{SO-HR CSA})}\left(\nicefrac{|\bm A|}{r}+|\bm B|,\ell+1,N,r\right)+\lambda_{\text{\textbullet}}rN|\bm B|.\label{eq:enc_complexity_SCSA_0}
\end{align} Naturally, if $|\bm A|>|\bm B|$ ($|\bm A|<|\bm B|$), $\text{SCSA}(0)$ ($\text{SCSA}(1)$) gives a lower encoding complexity than $\text{SCSA}(1)$ ($\text{SCSA}(0)$). For given matrices $\bm A$ and $\bm B$, we compute the SO-HR SCSA encoding complexity as
\begin{align}
	&C_{\text{EC}}^{\text{SO-HR SCSA}}=\min_{b\in\{0,1\}}C_{\text{EC}}^{\text{SO-HR SCSA}(b)}\nonumber\\
		&=C_{\text{EC}}^{(\text{SO-HR CSA})}\left(\nicefrac{\max\left(|\bm A|,|\bm B|\right)}{r}+\min\left(|\bm A|,|\bm B|\right),\ell+1,N,r\right)+\lambda_{\text{\textbullet}}rN\min\left(|\bm A|,|\bm B|\right),
\end{align} whereas the encoding complexity of HR SCSA is
\begin{align}
	C_{\text{EC}}^{\text{HR SCSA}}&=\sum_{i=1}^{N}\sum_{j=1}^{r}C_{\text{EC}}^{\text{HR}}\left(\ell\left(\nicefrac{\max\left(|\bm A|,|\bm B|\right)}{r}+\min\left(|\bm A|,|\bm B|\right)\right)\right)+\lambda_{\text{\textbullet}}rN\min\left(|\bm A|,|\bm B|\right)\nonumber\\
		&=Nr\ell\left(\nicefrac{\max\left(|\bm A|,|\bm B|\right)}{r}+\min\left(|\bm A|,|\bm B|\right)\right)+\lambda_{\text{\textbullet}}rN\min\left(|\bm A|,|\bm B|\right).
\end{align} Then, the theoretical gain of SO-HR SCSA over HR SCSA is given by
\begin{align}
	1-\frac{C_{\text{EC}}^{\text{SO-HR SCSA}}}{C_{\text{EC}}^{\text{HR SCSA}}}=\frac{\left(\ell-\lambda_{+}\right)\left(N-\bar{y}_D-1\right)\left(\nicefrac{\max\left(|\bm A|,|\bm B|\right)}{r}+\min\left(|\bm A|,|\bm B|\right)\right)}{C_{\text{EC}}^{\text{HR SCSA}}}.\label{eq:SCSA_enc_gain_theory}
\end{align} The complexity and the theoretical gains for SO-HR USCSA and GSCSA can be derived similarly. Details on the derivation are omitted here for the sake of brevity. The final expressions are provided in Table \ref{tab:enc_complexity_other CSA schemes}.

\begin{table}
	\caption{\small{Encoding complexity of USCSA and GSCSA}}
	\begin{center}
		\footnotesize{
			\begin{tabular}{ | c | c | c| }
				\hline
				\multirow{2}{*}{Scheme} & \multirow{2}{*}{Sub-Scheme} & \multirow{2}{*}{\shortstack{Complexity \\ $\left[\text{operation}\right]$}\footnote{We ignore the complexity of computing the scalars $\prod_{k=1}^{g}(k+(j-1)g+\alpha_i)$, $\forall i\in[1:N]$, $\forall j\in[1:\bar{g}]$.}} \\ 
				& & \\ \hline
				\multirow{2}{*}{USCSA} & $\text{USCSA}\left(f,q,g,0\right)$ & $C_{\text{EC}}^{(\text{SO-HR CSA})}\left(\frac{|\bm A|}{g}+\frac{|\bm B|}{\bar{g}},\ell,N,\bar{g}\right)+N\left(\bar{g}|\bm A|+|\bm B|\right)$ \\ \cline{2-3}
				& $\text{USCSA}\left(f,q,g,1\right)$ & $C_{\text{EC}}^{(\text{SO-HR CSA})}\left(\frac{|\bm A|}{\bar{g}}+\frac{|\bm B|}{g},\ell,N,\bar{g}\right)+N\left(|\bm A|+\bar{g}|\bm B|\right)$ \\ \hline
				\multirow{2}{*}{GSCSA} & $\text{GSCSA}\left(f,q,g,0\right)$ & $C_{\text{EC}}^{(\text{SO-HR CSA})}\left(\frac{|\bm A|}{fq}+|\bm B|,\ell,N,\bar{g}\right)+N\left(|\bm A|+\bar{g}|\bm B|\right)$ \\ \cline{2-3}
				& $\text{GSCSA}\left(f,q,g,1\right)$ & $C_{\text{EC}}^{(\text{SO-HR CSA})}\left(|\bm A|+\frac{|\bm B|}{fq},\ell,N,\bar{g}\right)+N\left(\bar{g}|\bm A|+|\bm B|\right)$ \\ \hline 
			\end{tabular}}
	\end{center}
	\label{tab:enc_complexity_other CSA schemes}
\end{table}
          
\section{Python Implementation}

Five schemes (SCSA, USCSA, GSCSA, A3S and GASP) have been implemented using the python library \textit{MPI4py} that provides the Message Passing Interface (MPI). The implementation is available on GitHub \cite{py_implementation}. In order to create a cluster we have used StarCluster, an open source tool for cluster computing \cite{starcluster}. We created a cluster on Amazon EC2, consisting of $N+1$ hosts -- one user and $N$ servers, which run on the instances \verb|c3.8xlarge|. To compare the schemes, we measure the time for $(a)$ computing the encoding matrices $\left(\tilde{\bm A}_i,\tilde{\bm B}_i\right)_{i=1}^{N}$ ($T_{\text{EC}}$), $(b)$ the upload of the encoding matrices ($T_{\text{UL}}$), $(c)$ the server computation $(T_{\text{C}})$, $(d)$ the download of $\left(\bm O_i\right)_{i\in\mathcal{Q}}$ $(T_{\text{DL}})$ and $(e)$ the decoding of $\bm A\bm B$ $(T_{\text{DC}})$. As far as the computation at the servers is concerned, we do not measure the whole time that is spent by all servers to compute $\bm O_i$ but rather the average time $\overline{T}_{\text{C}}$ over the number servers. However, because all servers exhibit very similar timing overhead for the computation, the fluctuation around the mean value is very small and can therefore be neglected. 
\begin{figure}
	\centering
	\begin{tikzpicture}[roundnode/.style={circle, draw=green!60, fill=green!5, very thick, minimum size=7mm}, squarednode/.style={rectangle, draw=red!60, fill=red!5, very thick, minimum size=5mm, rounded corners},scale=0.9]
	
		\draw [->,line width=1.25,>=stealth] (0,0) -- (12,0);
		\draw [line width=1.25,>=stealth] (0,0) -- (0,7);
		\draw [dashed] (0,1.5) -- (12,1.5);
		\node[draw, rotate=90, anchor=east, draw=white] at (-0.5,1.35) {\small{\textbf{User}}};
		\node[draw, rotate=90, anchor=east, draw=white] at (-0.5,4.5) {\small{\textbf{Server}}};
		\node[draw, anchor=east, draw=white] at (13.25,0) {\small{Time}};
		
		\node[squarednode,fill=dandelion!30, draw=dandelion, anchor=south west] (enc) at (0.025, 0.5) {\footnotesize{$\quad\text{User}\quad$}};
		\draw[<->, thick, draw=dandelion, thick,>=stealth] (0.025,6) -- (1.8,6) node[pos=0.5,sloped,above, thick] {\footnotesize $T_{\text{EC}}$};
		\node[draw, anchor=center, draw=white, minimum size=2mm, align=center] at (1.8,-0.6) {\footnotesize{UL-Tx} \\[-2ex] \footnotesize{starts}};
		\draw [dashed] (1.8,0) -- (1.8,7);
		
		\node[squarednode,fill=green!30, draw=OliveGreen, anchor=south west] (u1) at (1.82, 4.5) {\footnotesize{$\quad\text{Server }1\quad$}};
		\node[squarednode,fill=green!30, draw=OliveGreen, anchor=north west] (u2) at (2.3, 4.45) {\footnotesize{$\quad\text{Server }2\quad$}};
		\node[squarednode,fill=green!30, draw=OliveGreen, anchor=north west, align=center] (u3) at (2.1, 3.8) {\footnotesize{$\quad\quad\text{Server }3\quad\quad$}};
		\node[draw, anchor=east, draw=white, minimum size=2mm] at (3.7,2.7) {{$\vdots$}};
		\node[squarednode,fill=green!30, draw=OliveGreen, anchor=south west, align=center] (uN) at (2.1, 1.51) {\footnotesize{$\quad\:\text{Server }N\quad\:$}};
		\draw [dashed] (5.2,0) -- (5.2,7);
		\draw[<->, thick, draw=OliveGreen, thick,>=stealth] (1.825,5.5) -- (5.2,5.5) node[pos=0.5,sloped,above, thick] {\footnotesize $T_{\text{UL}}$};

		\node[squarednode,fill=blue-violet!30, draw=blue-violet, anchor=south west] (c1) at (4.18, 4.5) {\footnotesize{$\quad\text{Server }1\quad$}};
		\node[squarednode,fill=blue-violet!30, draw=blue-violet, anchor=north west] (c2) at (4.65, 4.45) {\footnotesize{$\qquad\text{Server }2\qquad$}};
		\node[squarednode,fill=blue-violet!30, draw=blue-violet, anchor=north west, align=center] (c3) at (5.2, 3.8) {\footnotesize{$\text{Server }3$}};
		\node[draw, anchor=east, draw=white, minimum size=2mm] at (6.2,2.7) {{$\vdots$}};
		\node[squarednode,fill=blue-violet!30, draw=blue-violet, anchor=south west, align=center] (cN) at (4.71, 1.51) {\footnotesize{$\:\text{Server }N\:$}};
		\draw [dashed] (7.74,0) -- (7.74,7);
		\draw[<->, thick, draw=blue-violet, thick,>=stealth] (5.2,6) -- (7,6) node[pos=0.5,sloped,above, thick] {\footnotesize $\bar{T}_{\text{C}}$};
		\node[draw, anchor=center, draw=white, minimum size=2mm, align=center] at (7.75,-0.6) {\footnotesize{DL-Tx} \\[-2ex] \footnotesize{starts}};
		
		\node[squarednode,fill=carrotorange!30, draw=carrotorange, anchor=south west] (d1) at (7.76, 4.5) {\footnotesize{$\text{Server }1$}};
		\node[squarednode,fill=carrotorange!30, draw=carrotorange, anchor=north west] (d2) at (7.76, 4.45) {\footnotesize{$\:\text{Server }2\:$}};
		\node[squarednode,fill=carrotorange!30, draw=carrotorange, anchor=north west, align=center] (d3) at (7.76, 3.8) {\footnotesize{$\:\:\text{Server }3\:\:$}};
		\node[draw, anchor=east, draw=white, minimum size=2mm] at (8.9,2.7) {{$\vdots$}};
		\node[squarednode,fill=carrotorange!30, draw=carrotorange, anchor=south west, align=center] (dN) at (7.76, 1.51) {\footnotesize{$\:\text{Server }N\:$}};
		\draw [dashed] (9.74,0) -- (9.74,7);
		\draw[<->, thick, draw=carrotorange, thick,>=stealth] (7.76,5.5) -- (9.74,5.5) node[pos=0.5,sloped,above, thick] {\footnotesize $T_{\text{DL}}$};
		\node[draw, anchor=center, draw=white, minimum size=2mm, align=center] at (9.74,-0.6) {\footnotesize{Decoding} \\[-2ex] \footnotesize{starts}};
		
		\node[squarednode,fill=chocolate(traditional)!30, draw=chocolate(traditional), anchor=south west] (enc) at (9.75, 0.5) {\footnotesize{$\quad\text{User}\quad$}};
		\draw [dashed] (11.55,0) -- (11.55,7);
		\draw[<->, thick, draw=chocolate(traditional), thick,>=stealth] (9.75,6) -- (11.55,6) node[pos=0.5,sloped,above, thick] {\footnotesize $T_{\text{DC}}$};
	
	\end{tikzpicture}
	\caption{\small{Time measurement setup in our implementation. In the figure, we assume that the recovery threshold matches the number of servers $N$.}}
	\label{fig:time_diagramm}
\end{figure}
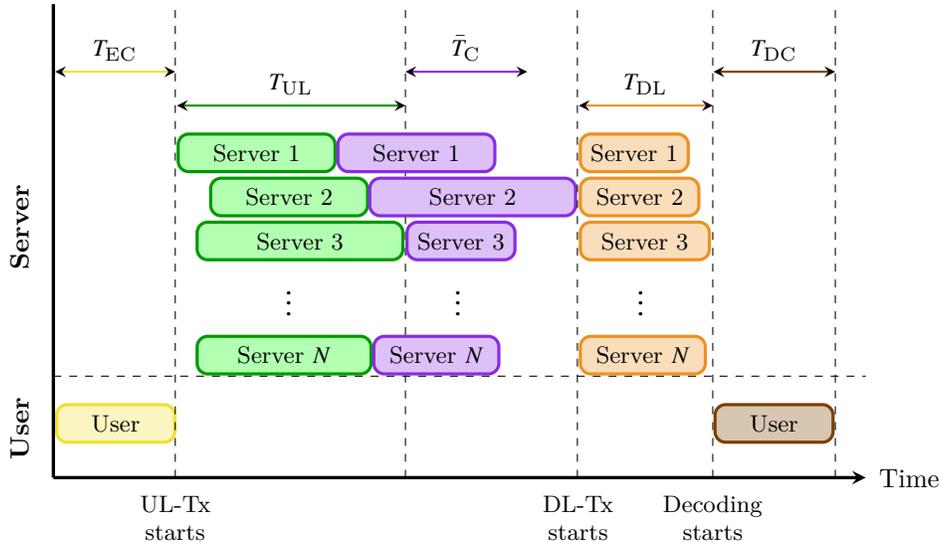

The total time, which was spent on the SDMM process is computed as the sum (cf. Fig. \ref{fig:time_diagramm})
\begin{align}
	\hat{T}=T_{\text{EC}}+T_{\text{UL}}+\overline{T}_{\text{C}}+T_{\text{DL}}+T_{\text{DC}}.\label{eq:time}
\end{align} 

\subsection{Numerical Results}

We test our implementation for two scenarios. In each scenario, we gradually increase the matrix dimensions of $\bm A$ and $\bm B$ -- namely $m$, $n$ and $p$ -- according to the sequence
\begin{align*}
	c_{j}&=1.3c_{j-1},\forall c\in\left\{m,n,p\right\},\forall j\in[1:9]
\end{align*} and measure the times $T_{\text{EC}}$, $T_{\text{UL}}$, $\bar{T}_{\text{C}}$, $T_{\text{DL}}$, $T_{\text{DC}}$ and $\hat{T}$ for $(m_k,n_k,p_k)$, $\forall k\in[0:9]$. We average each time measurement over $10$ iterations and plot the time measurements over a single matrix dimension -- namely $p$. Next, we specify the two considered scenarios.

\subsubsection*{Scenario $1$} In this scenario, we choose $\ell=4$, $N=15$, $m_0=90$, $p_0=1000$ and $n_0\in\{10,100\}$. We call Scenario $1$ with $n_0=10$ ($n_0=100$) the small-$n$ (big-$n$) scenario. All schemes apply a $\mathsf{PART}\left([r_{\bm A}, 1],[1,r_{\bm B}]\right)$ matrix partitioning scheme. The schemes are parametrized as follows:
\begin{itemize}
	\item A3S: $\left(r_{\bm A}^{\text{A3S}},r_{\bm B}^{\text{A3S}}\right)=\left(1,4\right)$,
	\item GASP: $\left(r_{\bm A}^{\text{GASP}},r_{\bm B}^{\text{GASP}}\right)=\left(2,2\right)$,
	\item SCSA: $\left(r_{\bm A}^{\text{SCSA}},r_{\bm B}^{\text{SCSA}}\right)=\left(1,7\right)$,
	\item USCSA: $\left(r_{\bm A}^{\text{USCSA}},r_{\bm B}^{\text{USCSA}}\right)=\left(2,3\right)$,
	\item GSCSA: $\left(r_{\bm A}^{\text{GSCSA}},r_{\bm B}^{\text{GSCSA}}\right)=\left(1,6\right)$
\end{itemize} and $f=2$ and $q=3$ for USCSA and GSCSA. Note that in the encoding process, we apply SO-HR for A3S, SCSA, USCSA and GSCSA and in the computation process we use a hybrid Winograd-Strassen (WS) algorithm proposed in \cite{Huss-lederman} for matrix multiplication in SCSA. The results for this scenario are shown in Fig. \ref{fig:results224_small_n} ($n_0=10$) and in Fig. \ref{fig:results224_big_n} ($n_0=100$).  

\subsubsection*{Scenario $2$} In this scenario, we set $\ell=2$, and $N=18$ and keep $m_0$, $p_0$ and $n_0$ as in Scenario $1$. Since $\ell$ and $N$ changed (as compared to Scenario $1$), we adapt the schemes according to:
\begin{itemize}
	\item A3S: $\left(r_{\bm A}^{\text{A3S}},r_{\bm B}^{\text{A3S}}\right)=\left(2,4\right)$,
	\item GASP: $\left(r_{\bm A}^{\text{GASP}},r_{\bm B}^{\text{GASP}}\right)=\left(3,3\right)$,
	\item SCSA: $\left(r_{\bm A}^{\text{SCSA}},r_{\bm B}^{\text{SCSA}}\right)=\left(1,14\right)$,
	\item USCSA: $\left(r_{\bm A}^{\text{USCSA}},r_{\bm B}^{\text{USCSA}}\right)=\left(3,4\right)$,
	\item GSCSA: $\left(r_{\bm A}^{\text{GSCSA}},r_{\bm B}^{\text{GSCSA}}\right)=\left(1,12\right)$				
\end{itemize} and $f=3$ and $q=4$ for USCSA and GSCSA. We plot the results for this scenario in Fig. \ref{fig:results332_small_n} for $n_0=10$ and in Fig. \ref{fig:results332_big_n} for $n_0=100$. Now we discuss the numerical results. We discuss each time measurement individually in the following paragraphs.

\subsubsection{Encoding}

\begin{figure}[h]
	\centering
	\resizebox{.8\linewidth}{!}{
		\input{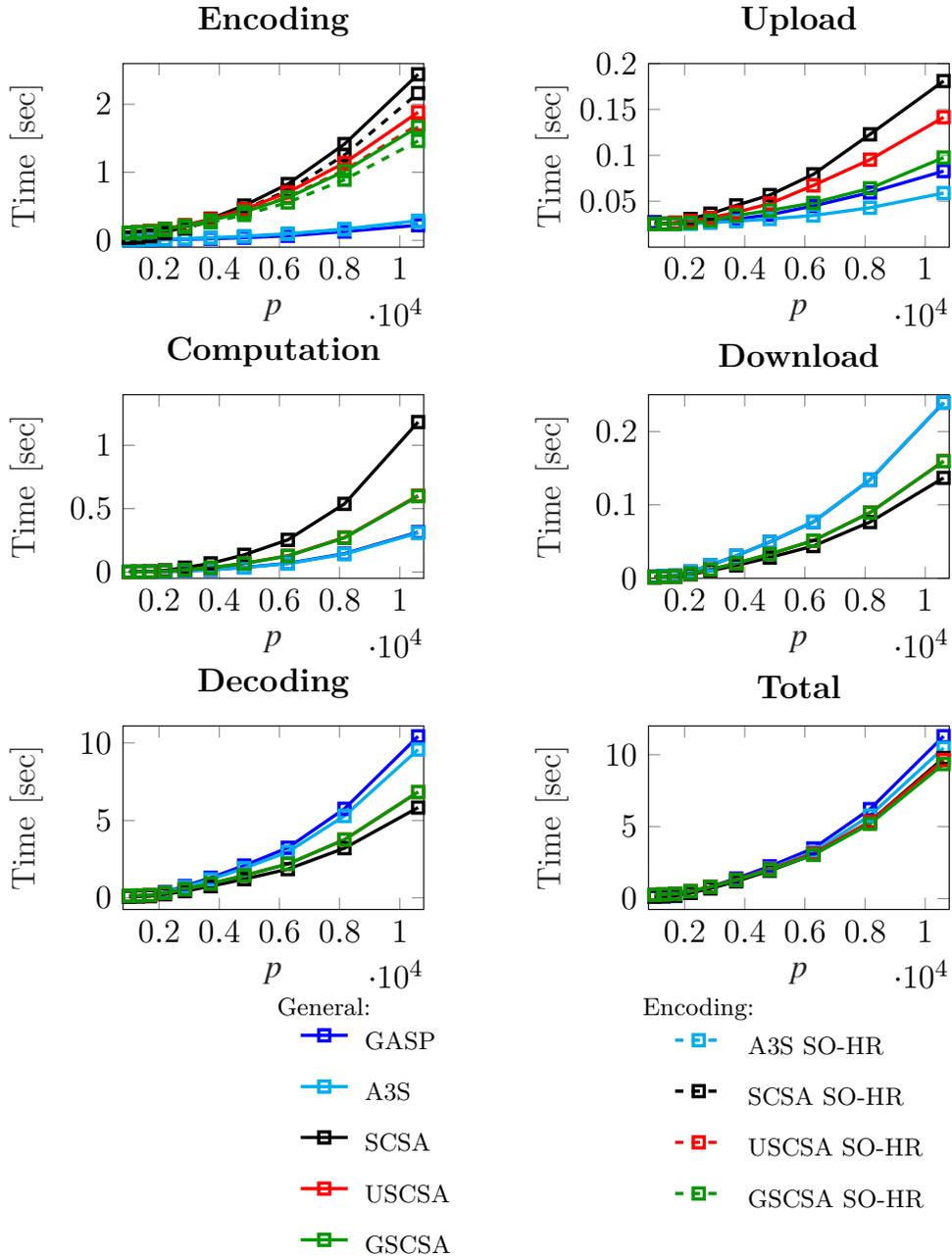}
	}
	\caption{\small{Scenario $1$ -- small-$n$ ($n_0=10$). Note that there are two legends - a general and an encoding-specific legend.}}
	\label{fig:results224_small_n}
\end{figure}

\begin{figure}[h]
	\centering
	\resizebox{.8\linewidth}{!}{
		\input{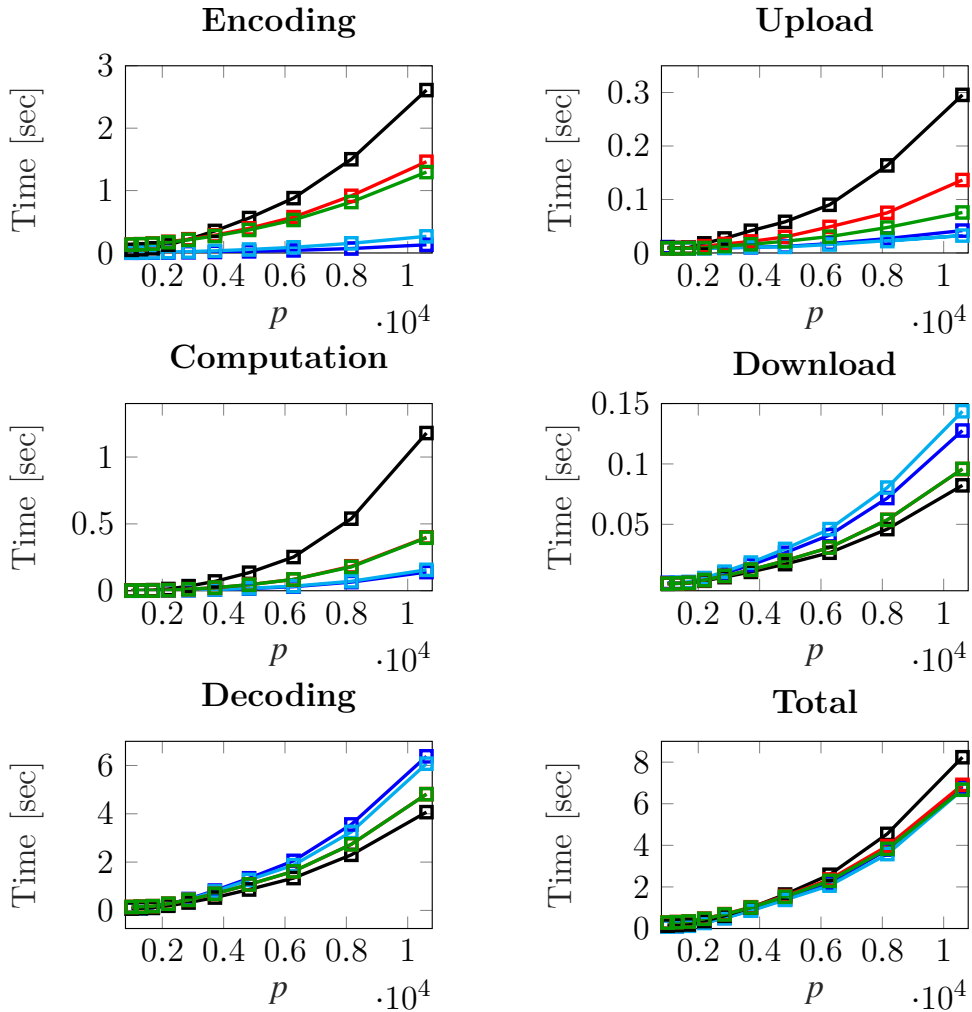}
	}
	\caption{\small{Scenario $2$ -- small-$n$ ($n_0=10$).}}
	\label{fig:results332_small_n}
\end{figure}

The encoding time is the duration needed to compute all encoding matrices $\left(\tilde{\bm A}_i,\tilde{\bm B}_i\right)_{i=1}^{N}$. While in cross subspace alignment (CSA) schemes -- SCSA, USCSA and GSCSA -- the encoding matrices for server $i$ consist of $\bar{j}$ pairs $\left(\tilde{\bm A}_i^{(j)},\tilde{\bm B}_i^{(j)}\right)_{j=1}^{\bar{j}}$, the encoding in GASP and 
A3S is lighter in the sense that only a single encoding pair ($\bar{j}=1$) is needed. 
Overall, we observe that GASP and A3S outperform all CSA schemes (cf. \ref{fig:results224_small_n}-\ref{fig:results332_big_n}). Our results also show that SO-HR represents a viable option to reduce the encoding complexity. They are particularly relevant for CSA schemes, where the absolute time saving over HR is significant. For instance, for Scenario $1$ with $n_0=10$ at $p\approx 1.6\cdot 10^{4}$, SO-HR SCSA gives a relative encoding gain of $10.9\%$ which closely matches the theoretical gain of Eq. \eqref{eq:SCSA_enc_gain_theory} ($10.8\%$). By comparing the performance of SO-HR for CSA schemes in Scenario $1$ ($\ell=4$) with Scenario $2$ ($\ell=2$), we could also verify that the SO-HR encoding gain decays with decreasing $\ell$. In the two scenarios considered in this paper, where the degree of the polynomials are relatively small, we observed that the presence of HR in conjuction with the second-order (SO) scheme is of negligible importance. However, we expect that as we increase the degree of the polynomial the influence of Horner's rule in reducing the encoding time will become more dominant. 

Alternative algorithms - amongst others the finite field fast fourier transform (FFT) \cite{Cooley_1965, Pollard_1971, Moenck_1976} -- have been tested in the encoding for the evaluation of matrix polynomials. However, they did not show good performance results for the two experimentation scenarios. The reasons are the following. For sparse matrix polynomials (cf. with A3S encoding) and low-degree matrix polynomials (cf. with CSA encoding), the FFT is not well suited, as the operation count, thus ultimately the running time, is dependent upon the degree $d-1$ of the matrix polynomial $\bm U^{(j)}(\alpha)$. Other issues that involve the finite field FFT are the power-of-two restriction on $d$ and the choice of a (primitive) root of unity for a given field. All these issues make the FFT not useful for the two experimentation scenarios. Alternative algorithms via division \cite{Moenck_1972} have also been tested for encoding but showed inferior performance to SO-HR.         

\subsubsection{Upload}

This time is devoted to send the encoded matrices $\tilde{\bm A}_i$ and $\tilde{\bm B}_i$ to servers $i\in[1:N]$. It is easy to see that for a sufficiently large $p$, A3S outperforms all remaining schemes in all scenarios. Further, we see that in comparison to SCSA both GSCSA and USCSA reduce the upload time significantly, e.g., for $p\approx 0.8\cdot 10^{4}$ the SCSA upload time is about $66\%$ ($47\%$) higher than of GSCSAs (USCSAs) upload time for Scenario $2$ with $n_0=100$. The dominance of USCSA and, in particular, GSCSA over SCSA is exacerbated with increasing matrix dimension $n$ (cf. upload time of Figs. \ref{fig:results332_small_n} and \ref{fig:results332_big_n}). Overall, the order in incurred upload time of all five schemes is in agreement with their order of upload costs.   

\subsubsection{Computation}

This time accounts for the computation of $\tilde{\bm A}_i$ and $\tilde{\bm B}_i$ at each server $i$. In schemes that use polynomial codes (PC), such as A3S and GASP, each server $i$ simply multiplies $\tilde{\bm A}_i\in\mathbb{F}^{\frac{m}{r_{\bm A}}\times n}$ with $\tilde{\bm B}_i\in\mathbb{F}^{n\times\frac{p}{r_{\bm B}}}$. In contrast to PC schemes, in CSA schemes such as SCSA, USCSA and GSCSA, each server $i$ multiplies $\tilde{\bm A}_{i}^{(j)}\in\mathbb{F}^{\frac{m}{r_{\bm A}}\times n}$ with its respective pair $\tilde{\bm B}_{i}^{(j)}\in\mathbb{F}^{n\times\frac{p}{r_{\bm B}}}$ and accumulates all $\bar{j}$\footnote{In our experiments, we set $\bar{j}$ to $\bar{j}=N-2\ell$ for SCSA and $\bar{j}=q$ for GSCSA and USCSA.} matrix products according to $\sum_{j=1}^{\bar{j}}\tilde{\bm A}_{i}^{(j)}\tilde{\bm B}_{i}^{(j)}$. For the scenarios under study, this results in the highest computation time for SCSA and the slowest for A3S and GASP. As expected, as we increase $n_0$ from $10$ to $100$, only the absolute time for the computation increases without affecting the relative order of the schemes. Advanced matrix multiplication algorithms developed by Strassen and extended by Winograd \cite{Strassen_1969,Coppersmith_1990} intertwined with regular matrix multiplication \cite{Huss-lederman}, show only improvements over pure regular matrix multiplication for the big-$n$ scenario when SCSA is applied. The gain at $p\approx 1.6\cdot 10^{4}$ for scenarios $1$ and $2$ are, respectively, $15.8\%$ and $15.6\%$.           

\subsubsection{Download}

\begin{figure}[h]
	\centering
	\resizebox{.8\linewidth}{!}{
		\input{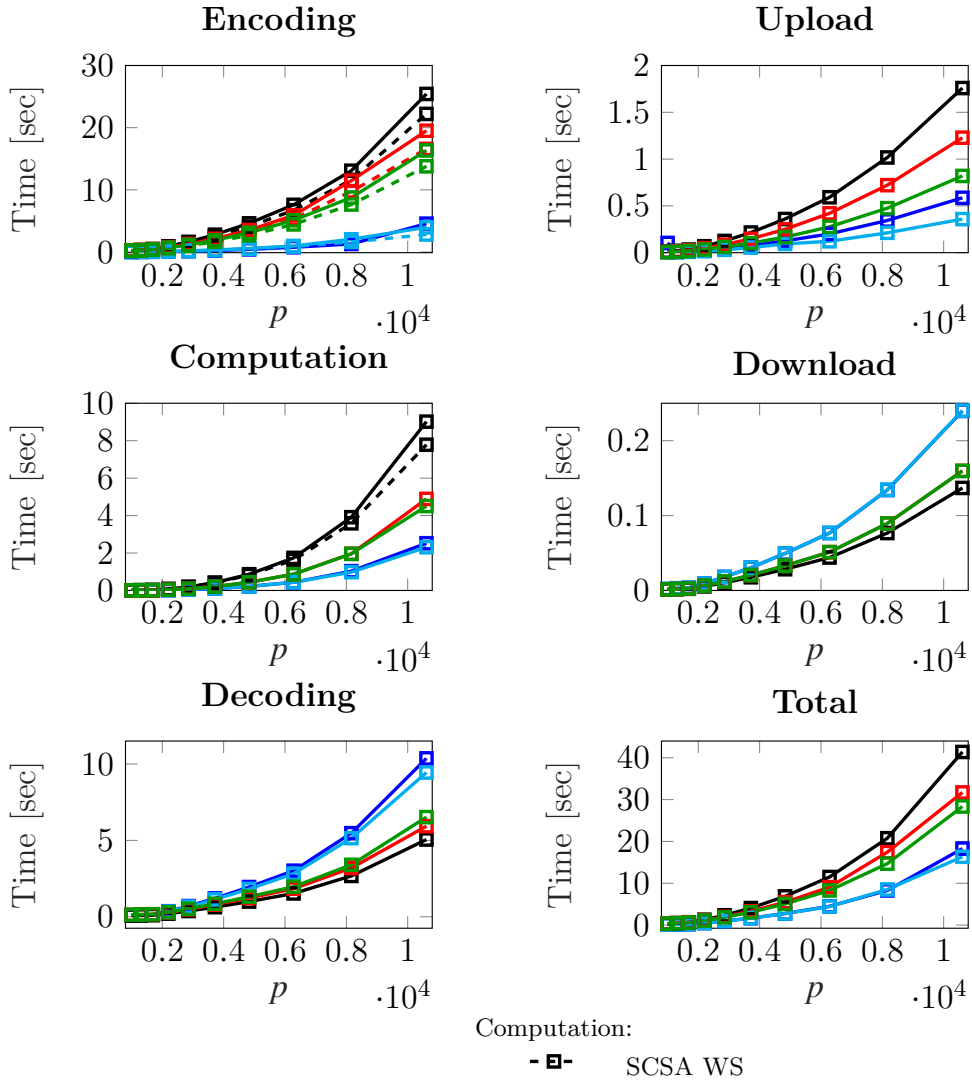}
	}
	\caption{\small{Scenario $1$ -- big-$n$ ($n_0=100$).}}
	\label{fig:results224_big_n}
\end{figure}

The download time in our experiment spans the duration until the user receives all $N$ server observations $\bm O_1,\bm O_2,\ldots,\bm O_N\in\mathbb{F}^{\frac{m}{r_{\bm A}}\times\frac{p}{r_{\bm B}}}$\footnote{In our experiments, the recovery threshold $Q$ matches the number of servers $N$.}. Since $K_{\text{DL}}^{\text{SCSA}}\leq K_{\text{DL}}^{\text{G(U)SCSA}}\leq K_{\text{DL}}^{\text{GASP}}\leq K_{\text{DL}}^{\text{A3S}}$ for a constant $\ell$ and $N$, we expect that for sufficiently large matrix dimensions SCSA will outperform all remaining schemes with respect to the download time. Our numerical results confirm this expectation. Further, the numerical results also verify that for all schemes the download time does not depend on the matrix dimension $n$ (cf. download time of Figs. \ref{fig:results224_small_n} and \ref{fig:results224_big_n}).          

\subsubsection{Decoding}

The decoding time encompasses the time at the user $(i)$ to construct the $N\times N$ decoding matrix $\bm D$, $(ii)$ to determine the inverse $\bm D^{-1}$ and $(iii)$ to multiply the inverse $\bm D^{-1}$ with the set of $N$ server observations $\bm O_{[1:N]}$. For the inversion we use python built-in functions instead of fast algorithms for the inversion of Vandermonde matrices \cite{Pollard_1971,Gao_2002,Kedlaya_2011} in case of polynomial codes or Cauchy-Vandermonde matrices \cite{Finck_1993} for CSA codes. Overall, the time for steps $(i)$ and $(ii)$ are (a) independent of the matrix dimensions $m,n$ and $p$ and (b) behave similarly for all five schemes. On the contrary, the time for step $(iii)$ scales with $\frac{1}{r_{\bm A}r_{\bm B}}$ and $mp$ and thus vary for different schemes. Again, the numerical results validate this behavior for large matrix dimensions. For instance, SCSA has the largest $r_{\bm A}r_{\bm B}$-value and simultaneously has the lowest decoding time.     
\subsubsection{Total Time}

\begin{figure}[h]
	\centering
	\resizebox{.8\linewidth}{!}{
		\input{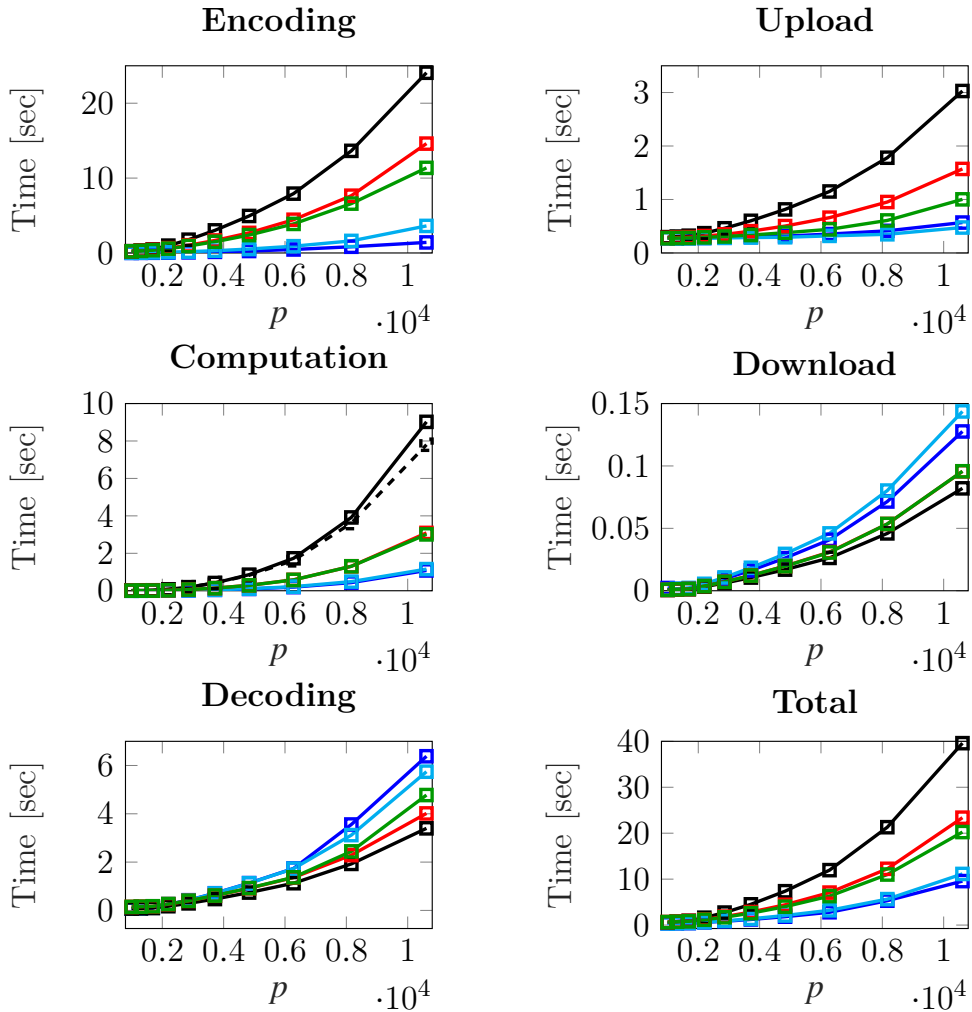}
	}
	\caption{\small{Scenario $2$ -- big-$n$ ($n_0=100$).}}
	\label{fig:results332_big_n}
\end{figure}

We compute the total time according to Eq. \eqref{eq:time} and plot the results in Figs. \ref{fig:results224_small_n}--\ref{fig:results332_big_n}.  While $n_0=100$ is a preferable situation for A3S and GASP (see Figs. \ref{fig:results224_big_n} and \ref{fig:results332_big_n}) over SCSA, SCSA tends to outperform GASP and A3S in situations where $n_0=10$. GSCSA and USCSA, on the other hand, possess good performance results over both scenarios irrespective of the $n_0$ value. This is mainly due to the fact that they combine advantages of GASP and A3S (i.e., upload and computation) with those of SCSA (i.e., download and decoding). 

\section{Conclusion}

In this paper, we studied the secure distributed matrix multiplication problem (SDMM), where a user is interested in computing the matrix product $\bm A\bm B$ of two private matrices $\bm A$ and $\bm B$. To this end, the user outsources the multiplication job to $N$ distributed servers without leaking any information to any set of $\ell\leq N$ colluding servers. In SDMM, the goal is to find schemes that optimally balance conflicting communication and computation metrics. One such tradeoff is the one between uplink and downlink communication efficiency. As part of this study, we first propose two uplink adjustable secure cross subspace alignment (CSA) schemes, namely USCSA and GSCSA that balance uplink and downlink communication costs. For CSA schemes, we develop a second-order (SO) encoding scheme that exploits on the symmetry property of evaluating a polynomial $u(x)$ at $x$ and $-x$. Next, we implement various SDMM schemes of the literature (including USCSA and GSCSA) in Python and compare computation and communication times using Amazon EC2 instances. Our numerical results show that USCSA and GSCSA establish a good balance between the time spend on the communication and computation in SDMMs. Further, we infer from our simulations that SO encoding represents a good solution to reduce the encoding complexity of CSA schemes. A future research thrust is to refine our schemes by using optimized decoding algorithms \cite{Gerasoulis_1987,Finck_1993}.                

\appendices

\section{Proof of Lemma \ref{lem:conv_K_UL}}
	 	Secure computing of $\bm A\bm B$ requires that 
	 	\begin{itemize}
	 		\item[(i)] $H(\bm A,\bm B|\tilde{\bm A}_{[1:N]},\tilde{\bm B}_{[1:N]})=0$ (cf. decodability constraint \eqref{eq:decod_constraint}) and
	 		\item[(ii)] $H(\bm A,\bm B)=H(\bm A,\bm B|\tilde{\bm A}_{\mathcal{L}},\tilde{\bm B}_{\mathcal{L}})$ for any $\mathcal{L}\subseteq[N],|\mathcal{L}|=\ell$ (cf. security constraint \eqref{eq:sec_constraint}).
	 	\end{itemize} Thus, we infer that
	 	\begin{align*}
		 	H(\bm A,\bm B)&=I(\tilde{\bm A}_{\mathcal{L}^{C}},\tilde{\bm B}_{\mathcal{L}^{C}};\bm A,\bm B|\tilde{\bm A}_{\mathcal{L}},\tilde{\bm B}_{\mathcal{L}})\\
			 	&=H(\tilde{\bm A}_{\mathcal{L}^{C}},\tilde{\bm B}_{\mathcal{L}^{C}}|\tilde{\bm A}_{\mathcal{L}},\tilde{\bm B}_{\mathcal{L}})-H(\tilde{\bm A}_{\mathcal{L}^{C}},\tilde{\bm B}_{\mathcal{L}^{C}}|\tilde{\bm A}_{\mathcal{L}},\tilde{\bm B}_{\mathcal{L}},\bm A,\bm B),
	 	\end{align*} where $\mathcal{L}^{C}=[1:N]\setminus\mathcal{L}$. Ignoring the second term of above mutual information term gives the upper bound
	 	\begin{align}
		 	H(\bm A,\bm B)\leq H(\tilde{\bm A}_{\mathcal{L}^{C}},\tilde{\bm B}_{\mathcal{L}^{C}}|\tilde{\bm A}_{\mathcal{L}},\tilde{\bm B}_{\mathcal{L}})\label{eq:bound_1_intermediate_1}.
	 	\end{align} Since there are ${N \choose N-\ell}$ possible subsets $\mathcal{L}^{C}$ of non-colluding servers of size $N-\ell$, we can sum up \eqref{eq:bound_1_intermediate_1} and obtain
	 	\begin{align*}
		 	{N \choose N-\ell}H(&\bm A,\bm B)\leq\sum_{\substack{\mathcal{L}^{C}\subseteq[1:N]\\|\mathcal{L}^{C}|=N-\ell}} H(\tilde{\bm A}_{\mathcal{L}^{C}},\tilde{\bm B}_{\mathcal{L}^{C}}|\tilde{\bm A}_{\mathcal{L}},\tilde{\bm B}_{\mathcal{L}})\nonumber\\
			 	\iff H(\bm A,\bm B)&\leq\frac{N-\ell}{{N \choose N-\ell}}\sum_{\substack{\mathcal{L}^{C}\subseteq[1:N]\\|\mathcal{L}^{C}|=N-\ell}}\frac{H(\tilde{\bm A}_{\mathcal{L}^{C}},\tilde{\bm B}_{\mathcal{L}^{C}}|\tilde{\bm A}_{\mathcal{L}},\tilde{\bm B}_{\mathcal{L}})}{N-\ell}
	 	\end{align*}
	 	Now, we can apply Han's concentration inequality \cite{Cover_2006} on conditional entropies to get
	 	\begin{align*}
		 	H(\bm A,\bm B)&\leq\frac{(N-\ell)}{N}H(\tilde{\bm A}_{[1:N]},\tilde{\bm B}_{[1:N]})\leq\frac{(N-\ell)}{N}\sum_{n=1}^{N}H(\tilde{\bm A}_{n},\tilde{\bm B}_{n})\nonumber\\
		 	&\leq\frac{(N-\ell)}{N}\left(\sum_{n=1}^{N}H(\tilde{\bm A}_{n})+H(\tilde{\bm B}_{n})\right)\leq\frac{(N-\ell)}{N}\left(\sum_{n=1}^{N}|\tilde{\bm A}_{n}|+|\tilde{\bm B}_{n}|\right)\log\mathbb{F}. 
	 	\end{align*} For independent and uniformly distributed matrices $\bm A$ and $\bm B$, $H(\bm A,\bm B)=n(m+p)\log\mathbb{F}$. For this case, rearranging above inequality gives the bound \eqref{eq:bound_K_UL}.

\ifCLASSOPTIONcaptionsoff
  \newpage
\fi

\bibliographystyle{IEEEtran}
\bibliography{Citations}

\end{document}